\documentclass[aoas]{imsart}

\RequirePackage{amsthm,amsmath,amsfonts,amssymb}
\RequirePackage[authoryear]{natbib}
\RequirePackage[colorlinks,citecolor=blue,urlcolor=blue]{hyperref}
\RequirePackage{graphicx}

\RequirePackage{bm}
\RequirePackage{mathrsfs}
\RequirePackage[justification=centering]{subfig}
\RequirePackage{caption}
\RequirePackage{multirow}
\RequirePackage[dvipsnames,svgnames,x11names]{xcolor}
\RequirePackage[normalem]{ulem}
\RequirePackage[]{pdfpages}

\newcommand{\tr}{\text{tr}}
\newcommand{\DD}{\mathbf{D}}
\newcommand{\RR}{\mathbf{R}}
\newcommand{\KK}{\mathbf{K}}
\newcommand{\XX}{\mathbf{X}}
\newcommand{\YY}{\mathbf{Y}}
\newcommand{\UU}{\mathbf{U}}

\newcommand{\GG}{\mathbf{G}}
\newcommand{\FF}{\mathscr{F}}
\newcommand{\SSS}{\mathbf{S}}

\newcommand{\SSigma}{\mathbf{\Sigma}}

\graphicspath{{./Paper Figures/}}

\startlocaldefs
\theoremstyle{plain}

\theoremstyle{remark}

\endlocaldefs

\begin{document}

\begin{frontmatter}
\title{Sparse precision matrix estimation in phylogenetic trait evolution models}
\runtitle{Graphical Phylogenetic Trait Evolution Model}

\begin{aug}
\author[A]{\fnms{Felipe}~\snm{Grillo Pinheiro}\ead[label=e1]{grillopf@gmail.com}},
\author[A]{\fnms{Taiane}~\snm{Schaedler Prass}\ead[label=e2]{ taiane.prass@ufrgs.br}},
\author[B]{\fnms{Gabriel W}~\snm{Hassler}\ead[label=e3]{ghassler@ucla.edu}},
\author[B,C]{\fnms{Marc A}~\snm{Suchard}\ead[label=e4]{msuchard@ucla.edu}}
\and
\author[A]{\fnms{Gabriela}~\snm{Bettella Cybis}\ead[label=e5]{gabriela.cybis@ufrgs.br}}

\address[A]{Department of Statistics,
Universidade Federal do Rio Grande do Sul, Porto Alegre, Brazil\printead[presep={,\ }]{e1,e2,e5}}
\address[B]{Department of Computational Medicine, University of California, Los Angeles, United States\printead[presep={,\ }]{e3}}
\address[C]{Departments of Biostatistics and Human Genetics, University of California, Los Angeles, United States\printead[presep={,\ }]{e4}}
\end{aug}

\begin{abstract}
Phylogenetic trait evolution models allow for the estimation of evolutionary correlations between a set of traits observed in a sample of related organisms. By directly modeling the evolution of the traits along an estimable phylogenetic tree, the model's structure effectively controls for shared evolutionary history. 
In these models, relevant correlations are usually assessed through the high posterior density interval of their marginal distributions.
However, the selected correlations alone may not provide the full picture regarding trait relationships. Their association structure, expressed through a graph that encodes partial correlations,
can in contrast highlight sparsity patterns featuring direct associations between traits. 
In order to develop a model-based method to identify this association structure we explore the use of Gaussian graphical models (GGM) for covariance selection.
We model the precision matrix with a G-Wishart conjugate prior, which results in sparse precision estimates. Furthermore the model naturally allows for Bayes Factor tests of association between the traits, with no additional computation required. 
We evaluate our approach through Monte Carlo simulations and applications that examine the association structure and evolutionary correlations of phenotypic traits in Darwin's finches and genomic and phenotypic traits in prokaryotes. Our approach provides accurate graph estimates and lower errors for the precision and correlation parameter estimates, particularly for conditionally independent traits, which are the target for sparsity in GGMs. 
\end{abstract}


\end{frontmatter}


\section{Introduction}
\label{sec:Intro}

Estimating evolutionary correlations between a set of traits evolving along a phylogenetic tree has been the focus of recent models that aim to elucidate trait interconnection. Here we explicitly model the association structure through a graph parameter in a Bayesian phylogenetic trait evolution model, improving inference for precision and correlation parameters and our ability to unravel complex trait relationships.

Phylogenetic trait evolution (PTE) models  are important tools for investigating the evolutionary associations between a set of phenotypic and/or genotypic  traits while controlling for the shared evolutionary history of related organisms. Failing to control for this evolutionary history can mislead the inference of correlations between these traits, as some of the associations could be simply a consequence of common ancestry rather than having an specific adaptive meaning.
To jointly model trait evolution along an unknown tree, several versions of phylogenetic trait evolution models have been proposed over the last few years \citep{cybis2015, hassler2020, zhang2021, zhang2022} building upon the Brownian diffusion process and the threshold model proposed by \citet{felsenstein2012}. 

These models assume that the observed traits of interest are linked to unobserved continuous latent variables for each tip taxon that arise from a multivariate Brownian diffusion (MBD) process along a phylogenetic tree inferred from molecular sequences. This MBD model is characterized by a precision matrix, the inverse covariance $\KK=\SSigma^{-1}$, from which the evolutionary trait correlations are obtained. The diffusion correlation $\RR$ is the parameter of scientific interest and informs the correlation between latent parameters, which is a proxy for the desired correlation between observed phenotypic traits. The diffusion correlation can be viewed as the combined effect of relevant genetic factors (e.g.\ selective pressures, genetic linkage) that couples the evolution of observed traits after adjusting for the taxa shared evolutionary history.

The threshold model of \citet{felsenstein2012} adapted the MBD process to allow for evolutionary correlation estimation among binary and continuous phenotypic traits. The aim of \citet{cybis2015} was to bring the threshold model to a Bayesian perspective and extend it to create the phylogenetic multivariate latent liability model (PMLLM) for the evolution of mixed-type traits, accounting for continuous, binary, categorical, and ordinal data. 
The primary contribution of \citet{zhang2021} was to develop an efficient inference framework, called phylogenetic multivariate probit model (PMPM), based on the bouncy particle sampler (BPS), to sample the latent parameters from a high-dimensional truncated normal distribution, thus improving mixing and efficiency compared to the MCMC scheme in \citet{cybis2015}. 
Further exploring the \textit{Hamiltonian zigzag dynamics} (Zigzag-HMC), \citet{zhang2022} developed a joint sampling
scheme for highly correlated latent variables and correlation matrix elements taking advantage of improved mixing. Additionally, \citet{hassler2020} extended the sampling mechanism for the latent variables to provide a highly efficient likelihood computation under incomplete continuous trait data, allowing for correlation estimation in challenging scenarios. 

Despite the efforts to improve efficiency and expand the model's applicability, the significance of correlation coefficients is usually determined by the evaluation of their marginal posterior distribution using a high posterior density (HPD) interval. Selecting the relevant correlations is required because, in many problems, only a small portion of the observed traits are actually expected to be interconnected in the underlying biology. Thus, it is desirable to control for false positive signals and avoid estimation of spurious correlation coefficients, specially for high-dimensional problems where the number of traits $p$ is large. Moreover, one may want to favor a sparse representation of target parameters purely based on modeling motivations. 

In current practice, however, HPD is a post-processing procedure, such that relevant correlations are selected after parameters had been estimated. Nevertheless, a model-based approach that integrates correlation selection and parameter estimation might be a preferable strategy.  Additionally, the selected correlations alone may provide limited information regarding trait relationships.The partial correlations obtained from the diffusion precision parameter, in contrast, provide straightforward information about direct associations between traits, 
information hereafter referred to as association structure. When traits are conditionally independent --- i.e.\ not directly associated--- the precision matrix is likely to express some sparsity pattern. This sparsity reflects on the corresponding correlation matrix estimates potentially shrinking some of its off-diagonals towards zero, as desired. 
Therefore, estimating a sparse precision matrix is a natural systematic solution for performing covariance and correlation selection in PTE models. Notice that, although we refer to such as covariance or correlation selection, it is actually the precision matrix that has zero elements related to conditional independence which are being directly modeled \citep{gaskins2019}.

One way to obtain sparsity is by restricting the precision matrix to the space of positive definite matrices with zero entries consistent with a graph that also depicts the dependency structure between traits, a strategy used in covariance selection problems \citep{dempster1972}. For multivariate Gaussian variables with precision matrix $\KK=\{k_{ij}\}$, such as the latent parameters in phenotypic trait evolution models, the association structure can be translated by the conditional \allowbreak(in)dependence embedded in the precision matrix. Then an off-diagonal entry $k_{ij} = 0$ implies that the corresponding variables $(i,j)$ are conditionally independent given all the other variables in the model \citep{ li2020, mitsakakis2010, talhouk2012}. We refer the interested reader to \citet[Ch.9]{maathuis2018} for detailed proofs on the connection between conditional independence and the zero pattern in $\KK$.

Gaussian graphical models (GGM) are convenient tools for modeling conditional \allowbreak(in)dependence relationships among variables \citep{carvalho2009}. A GGM is a probabilistic model in which the conditional (in)dependence structure on $\KK$ is encoded in a graph $\GG$ \citep{atay2005, letac2007, mohammadi2015}. 
In this graph, variables are represented by the vertices and the presence or absence of edges indicates whether a direct association between them exists, such that non-zero entries in the off-diagonal of $\KK$ also correspond to the existing edges in the undirected graph $\GG$. 

In Bayesian inference, the G-Wishart distribution is the conjugate prior for structured precision matrices of a multivariate normal distribution. Because of its conjugacy, the G-Wishart is a convenient prior choice \citep{boom2021, williams2021} despite the lack of analytical form of its normalizing constant for a general non-decomposable graph $\GG$. The normalizing constant is required for G-Wishart density computations and can play an essential role in model selection --- the search for graphs (models) with high posterior density. For this reason, many different approaches have been proposed to compute or approximate the challenging $I_G(\delta,\DD)$. 
Some approaches focused on the advantages of estimating the normalizing constant for decomposable graphs \citep{letac2007} --- the constant has closed-form in this context ---, whereas others provide the pathway to its generalization to non-decomposable graphs \citep{roverato2000, roverato2002, atay2005}, further improving or avoiding its calculation \citep{lenkoski2011, letac2017,  mohammadi2015}, or developing closed-form expressions for specific graph configurations \citep{uhler2018}. 

In order to employ a model-based method to identify the association structure between traits and select the relevant correlations we explore the use of Gaussian graphical models.
In this study, we propose a Bayesian approach, called Graphical Phylogenetic Trait Evolution Model (GPTEM)
for inference of a sparse precision matrix $\KK$ by adapting the MBD model for continuous traits to the context of covariance and correlation selection. 
By estimating the association structure through the graph we aim to benefit from: 
i) a systematic solution for elimination of spurious correlations between phenotypic traits; 
ii) parameter reduction, which is a major gain since the number of pairwise correlations scales quadratically in trait dimension; 
iii) consequent error reduction in diffusion precision and correlation matrices estimation; and iv) improving our ability to explore complex relationships between continuous traits.

The reminder of this paper proceeds as follows: 
In Section \ref{sec: Modeling}, we first describe the MBD process on phylogenetic trees, further expand it to account for covariance selection using GGM, and define GPTEM.
In Section \ref{sec Inference}, we describe the inference framework for the MCMC and present a joint sampling scheme for the graph and sparse precision matrix. 
Section \ref{sec Simulation Study} presents the results from two simulation studies comparing the performance of GPTEM (also referred to as Graphical model) to  the traditional phylogenetic trait evolution model (Full model).
Then, we apply both graphical and full models to examine the association structure and evolutionary correlation of Darwin's finches phenotypic traits and prokaryote genomic and phenotypic growth properties in Section \ref{sec Application}. 
Finally, we 
discuss the advantages, limitations and future directions of GPTEM along with concluding remarks in Section \ref{sec: Discussion}.

\section{Modeling} 
\label{sec: Modeling}
In this section we show how we connect the continuous $p$-dimensional observed traits $\YY$ and correspondent latent variables $\XX$  to the phylogenetic tree $\FF$ using the multivariate Brownian diffusion model to control for shared evolutionary history of $N$ taxa in the estimation of trait correlations $\RR$. We explain how we adapt the MBD model to the context of covariance selection and present the Graphical Phylogenetic Trait Evolution Model. We also provide details on graph estimation (model selection) and sampling strategy for our Bayesian inference framework.

\subsection{Multivariate Brownian diffusion on Trees}
\label{subsec: MBD on Trees}
Consider a data set of $N$ aligned molecular sequences $\mathbf{S}$ from related organisms and $N$ continuous $p$-dimensional trait observations  $ \YY=  (\YY_1, \dots, \allowbreak \YY_N)^t$, where $\YY_k=(Y_{k1},\dots,Y_{kp})^t$ for $k=1,\dots,N$. 
We assume that $\YY$ arises from a partially observed multivariate Brownian diffusion process along a phylogenetic tree $\FF$. The tree $\FF = (\mathbb{V}, \mathbf{t})$ is a directed bifurcating acyclic graph with node set $\mathbb{V}$ and branch lengths $\mathbf{t}$. The node set $\mathbb{V}=(v_1,\dots, v_{2N-1})$ contains $N$ tip nodes of degree 1 $(v_1, \dots, v_N)$, $N-2$ internal nodes of degree 3 $(v_{N+1},\dots, v_{2N-2})$ and one root node $v_{2N-1}$ of degree 2. The branch lengths $\mathbf{t} = (t_1, \dots, t_{2N-2})$ denote the distance in real time from each node to its parent.

We associate each node $h$ in $\FF$ with a latent variable $\XX_h \in \mathbb{R}^p$ for $h = 1, \dots, 2N-1$. A multivariate Brownian diffusion process on $\FF$ characterizes the evolutionary relationship between latent variables $\XX$ and acts conditionally independently along each branch, such that $\XX_h$ is multivariate normally distributed,
\begin{equation}\label{eq: X_h}
    \XX_h \sim \mathcal{N}_{p}(\XX_{\text{pa}(h)}, t_h \KK^{-1}),
\end{equation}
centered at realized value $\XX_{\text{pa}(h)}$, where pa($h$) denotes the parent node of $h$, with variance proportional to a $p \times p$ positive definite covariance matrix $\KK^{-1}$ that is shared by all branches of $\FF$.
At the tips of $\FF$, we collect the $N \times p$ matrix $\XX = (\XX_1, \dots, \XX_N)^t$, where  $\XX_k = (X_{k1}, \dots, X_{kp})^t$ for $k=1,\dots,N$, and map it to the observed traits $\YY$ through a possibly stochastic link $p(\YY_k|\XX_k)$. The form of the stochastic link varies depending on the nature of $\YY$.
While the following approaches are valid for numerous choices of $p(\YY_k|\XX_k)$, we consider here only the case where the data $\YY$ are continuous and observed without measurement error or residual variance.
In this case, $p(\YY_k|\XX_k)$ has a degenerate density at $\XX_k$ (i.e.\ $\YY_k = \XX_k$ with probability 1).
We briefly discuss extensions to discrete or mixed-type traits, such as PMLLM of \citet{cybis2015} or PMPM of \citet{zhang2021} in Section \ref{sec: Discussion}.

Because we only need the latent parameters $\XX$ at the tips of $\FF$, i.e.\ $(\XX_1,\dots, \XX_N)^t$, to map $\YY$, we can compute the likelihood of $\XX$ by integrating out $\XX_{N+1},\dots ,\XX_{2N-1}$. In order to do so we adopt a conjugate prior on the root of the tree, $\XX_{2N-1} \sim \mathcal{N}_p(\bm\mu_0, \tau_0^{-1}\KK^{-1})$ with prior mean $\bm\mu_0$ and prior sample size $\tau_0$ \citep{pybus2012}. Hence, $\XX$ follows a matrix-normal distribution,
\begin{equation}\label{eq: MTN}
    \XX \sim \mathcal{MN}_{N\times p}(\mathbf{M},\mathbf{\Upsilon},\KK^{-1}),
\end{equation}
where $\mathbf{M} = \bm{1}_N \bm\mu_0^t$ is an $N \times p$ mean matrix, $\bm{1}_N$ is a vector of length $N$ populated by ones, $\KK^{-1}$ is the $p \times p$ across-trait covariance matrix, and $\mathbf{\Upsilon} = \mathbf{V}(\FF) + \tau_0^{-1} \mathbf{J}_N$ is an $N \times N$ across-taxa tree covariance matrix. The tree diffusion matrix $\mathbf{V}(\FF)$ is a deterministic function of $\FF$, and $\mathbf{J}_N=\bm1_N\bm1_N^t$ is an $N \times N$ matrix of all ones. The term $\tau_0^{-1} \mathbf{J}_N$ comes from the integrated-out tree root prior \citep[for further details on $\mathbf{V}$$(\FF)$, see][Figure 1]{zhang2021}.
Combining the stochastic link $p(\YY|\XX)$ and Equation (\ref{eq: MTN}) we can consider an augmented likelihood of $\YY$ and $\XX$ through the factorization
\begin{equation}\label{eq: likelihood}
    p(\YY,\XX |\KK,\FF, \bm\mu_0, \tau_0) = p(\YY|\XX)p(\XX | \KK,\FF, \bm\mu_0, \tau_0).
\end{equation}

Building upon the tree and matrix-normal structures, several algorithms were developed to evaluate the trait likelihood scaling linearly with the number of taxa in complete data scenarios \citep{tung2014, tolkoff2018}. When dealing with missing data, however, scalability becomes a bottleneck as the model requires data imputation or integration. 
To bypass this limitation, \citet{bastide2018} and \citet{hassler2020} propose an inference technique that integrates out missing values analytically and scales linearly with the number of taxa by using a post-order traversal algorithm.

In the following section we describe how we incorporate sparsity on the diffusion precision to perform covariance selection using GGM for this trait evolution model. Furthermore we present prior specifications for the parameters in the model. 

\subsection{Model extension: Covariance Selection with Gaussian Graphical Model} 
\label{subsec: model extension}
Gaussian graphical models provide a simple and convenient framework for imposing a conditional independence structure of association between variables \citep{mohammadi2015, williams2021}. Here we introduce some basic notation and the structure of undirected Gaussian graphical models. We refer the interested reader to \citet{lauritzen1996} for detailed information.

Let $\GG=(V,E)$ be an undirected graph defined by a finite set of vertices $V=\{1,2, \dots, p\}$ that represent the Gaussian variables, and a set of existing edges $E \subset \{(i,j)| 1 \leq i < j \leq p\}$ that represent links among the nodes $i,j \in V$. 
We define a $\bm{\mu}_0$ mean Gaussian graphical model with respect to the graph $\GG$ as the set of all Gaussian models such that
\begin{equation}\label{eq: M_G}
    \mathcal{M}_G = \{ \mathcal{N}_p(\bm{\mu}_0,\KK^{-1}), \KK \in \mathbb{P}_G \},
\end{equation}
where $\mathbb{P}_G$ is the space of $p \times p$ positive definite matrices with zero entries $(i, j)$ consistent with $\GG$, i.e.\ $k_{ij} = 0$ whenever $(i,j) \notin E$. Hence, in GGM, we assume that the precision matrix $\KK$ depends on the graph $\GG$.

\subsection{Graphical Phylogenetic Trait Evolution Model (GPTEM)}
\label{subsec: GPTEM}
Importing the GGM approach to the phylogenetic context we now restrict the precision matrix of the Brownian diffusion process to $\mathbb{P}_G$. This gives rise to the diffusion graph $\GG$, a novel parameter that represents the association structure of the partial correlations between the $p$ components of the latent variable $\XX$.  

We complete our model specification by choosing the prior distributions for the graph $\GG$ and for the structured precision matrix $\KK|\GG$.
For simplicity, we place a discrete uniform prior distribution over $\mathcal{G}$, the space of all graphs with $p$ edges \citep{mohammadi2015,mohammadi2021}, such that
\begin{equation}\label{eq: prior G}
    p(\GG) = \frac{1}{|\mathcal{G}|},
\end{equation}
for each $\GG \in \mathcal{G}$, where $|\mathcal{G}|=2^{p(p-1)/2}$ is the cardinality of the graph space. One can rather choose a different prior favoring denser or sparser graphs in the light of better knowledge about the association structure \citep{mohammadi2015}. 

For the prior distribution on the precision matrix $p(\KK|\GG)$, we use the G-Wishart distribution \citep{roverato2002, atay2005}, which is the conjugate prior, under Gaussian models, for structured precision matrices. The G-Wishart distribution $\mathcal{W}_G(\delta,\DD)$, with support on the space of $p \times p$ positive definite matrices, has density
\begin{equation}\label{eq:  prior K|G}
f_G(\KK;\delta,\DD) = \frac{1}{I_G(\delta,\DD)}| \KK | ^{(\delta - 2) / 2} \exp { \left\{ - \frac{1}{2} \tr( \DD\KK) \right\}}\bm{1}_{\KK \in \mathbb{P}_G},
\end{equation}
with parameters $\delta$ and $\DD$, where $\delta>0$ \citep[][Lemma 3.2.1: $I_G(\delta, \DD) < \infty$ for $\delta>0$]{mitsakakis2010} represents the degrees of freedom (or shape parameter), $\DD$ is a symmetric positive-definite rate matrix, and $I_G(\delta,\DD)$ is the normalizing constant, 
\begin{equation}\label{eq: W_G I_G}
     I_G(\delta, \DD)=\int_{\KK \in \mathbb{P}_G} | \KK | ^{(\delta - 2) / 2} \exp { \left\{ - \frac{1}{2} \tr( \DD\KK) \right\}} \,d\KK. 
\end{equation}
We use the Monte Carlo method of \citet{atay2005} to numerically approximate the prior and posterior normalizing constants required for graph updates during MCMC (for more detail, see Section A1 in Supplementary Material A \citep{pinheiro2022}).

\section{Inference} 
\label{sec Inference}
We single out the diffusion correlation $\RR$ and the diffusion graph $\GG$ as the primary parameters of scientific interest. The model is parametrized, however through the diffusion precision $\KK$ which indirectly shapes the correlations according to the conditional (in)dependencies in the graph.
Connecting the likelihood (\ref{eq: likelihood}) to the priors (\ref{eq: prior G}) and (\ref{eq:  prior K|G}), and dropping the posterior’s dependence on the hyperparameters $(\bm\Upsilon,\bm\mu_0, \tau_0 , \delta, \DD)$ to ease notation, we finally arrive at the posterior factorization
\begin{eqnarray}\label{eq: posterior}
 p(\KK, \GG, \FF | \YY, \textbf{S}) & \propto & p(\YY | \KK, \FF)  p(\KK|\GG)p(\GG) p(\textbf{S}|\FF )p(\FF) \nonumber \\
 & \quad = & \left( \int p(\YY| \XX) p(\XX|\KK, \FF) \,\text{d} \XX \right) p(\KK|\GG)p(\GG) p( \FF, \textbf{S}).
\end{eqnarray}

The joint posterior factorizes because sequences $\SSS$ only affect the parameters of primary interest through $\FF$, since we assume $\SSS$ to be conditionally independent of all other parameters given $\FF$ \citep{zhang2021}.
To obtain the posterior for this model we must integrate over the possible values for the unobserved latent variables at the tips of the tree.
For the simple MBD with continuous traits and no missing data, nevertheless, there is no need to perform integration over $\XX$ because of the nature of the link function.
To approximate the posterior distributions via MCMC simulation, we apply a random scan Metropolis-within-Gibbs \citep{liu1995} approach by which we sample parameter blocks one at a time at random, using Gibbs steps whenever a known full conditional distribution is available.  

\subsection{Sampling scheme}
\label{subsec: sampling scheme}
We employ standard Bayesian phylogenetic algorithms to obtain $p(\FF, \textbf{S})$ when the tree is unknown \citep{suchard2018}. Alternatively, $\FF$ can be fixed, in which case there is no need for sequence data $\SSS$.
If trait data $\YY$ is incomplete, we draw from the full conditional distribution of $\XX$ using the pre-order missing data augmentation algorithm developed by \citet{hassler2020} with overall computational complexity $\mathcal{O}(Np^3)$.

A joint updating scheme, inspired by the factorization $p(\GG,\KK)=p(\GG)p(\KK|\GG)$, is considered for $\KK$ and $\GG$. First, the graph $\GG$ is updated through a Metropolis-Hastings step whose target distribution is the marginal distribution of the graph. Then, $\KK$ is updated conditional on the new $\GG$ through a Gibbs step.

To update the graph $\GG$ we need to compute the marginal distribution of $\GG$, given all other parameters except $\KK$ (see Equation \eqref{SUPP: posterior G|X} in the \hyperref[SUPP: graph updates]{Appendix}). This marginal distribution is 
\begin{equation}\label{eq: G|X}
    p(\GG|\XX,\delta,\DD,\FF) \propto p(\GG|\delta,\DD)p(\XX|\GG,\delta,\DD,\FF)=\frac{p(\GG|\delta,\DD)}{(2\pi)^{Np/2}} \frac{I_G(\delta+N, \DD+\bm\Delta)}{I_G(\delta, \DD)},
\end{equation}
where $\bm \Delta =\left(\XX-\bm 1_N \bm\mu_0^t \right)^t  \left(\bm\Upsilon+\tau_0^{-1}\textbf{J}_N\right)^{-1}  \left(\XX-\bm{1}_N \bm\mu_0^t \right)$.
As in \citet{hassler2020}, we follow the methods in \citet{tung2014} to compute $\bm\Delta$ via post-order traversal of the tree, which has computational complexity $\mathcal{O}(Np^2)$. 

Based on this target distribution, the current graph $\GG$ is updated using a Metropolis-Hastings step with new graph proposals $\GG_p$ generated by switching the value of a randomly selected edge in $\GG$. 
We set $\GG=\GG_p$ by accepting the proposed graph with probability
\begin{equation}\label{eq: alpha}
    \alpha=\min{\left\{ 1,
    \frac{I_{G_p}(\delta+N, \DD+\bm\Delta)}{I_{G}(\delta+N, \DD+\bm\Delta)}\frac{I_{G}(\delta, \DD)}{I_{G_p}(\delta, \DD)}
     \right\}}.
\end{equation}
After the update, $\GG$ is then used to sample from the posterior distribution of $(\KK|\XX, \GG,\allowbreak \delta, \DD, \FF)$ in Equation (\ref{eq: posterior K|G}). 
Since we place a G-Wishart conjugate prior $\mathcal{W}_G(\delta,\DD)$ on the structured diffusion precision matrix, its full conditional distribution is also G-Wishart
\begin{equation}\label{eq: posterior K|G}
     \KK|\XX, \GG, \delta, \DD, \FF  \sim \\ \mathcal{W}_G\left(\delta+N, \DD+\bm\Delta \right).
\end{equation}

\subsection{Parameter Estimates}
After running the MCMC we process the posterior distributions of the parameters of scientific interest, namely $\GG$, $\KK$ and $\RR$. 
We estimate the entries of the posterior graph $\hat{\GG}=\{\hat{g}_{ij}\}$ from its MCMC iterations, after warm-up, through
\begin{equation}\label{eq: BF threshold}
    \hat{g}_{ij}=
        \begin{cases}
            1, & \text{if } BF_{1:0}^{(ij)} \geq 10^{1/2}\\
            0, & \text{otherwise}
        \end{cases},
\end{equation}
where $BF_{1:0}^{(ij)}$ is the Bayes factor calculated to evaluate the evidence in favor of including edge $(i,j)$ in the posterior graph estimate $\hat\GG$, i.e.\ evidence for conditional dependence between $i$-th and $j$-th traits. A BF above the threshold $10^{1/2}$ indicates substantial evidence for the hypothesis according to the criteria in \citet{jeffreys1998}.
We compute the Bayes factor as 
\begin{equation}\label{eq: BF}
  BF_{1:0}^{(ij)}=\frac{ p(\hat{g}_{ij}=1|\YY,\XX) }{ p(\hat{g}_{ij}=0|\YY,\XX ) } \frac{p(\hat{g}_{ij}=0) }{ p(\hat{g}_{ij}=1) } = \frac{ \hat{pe}_{ij} }{ (1-\hat{pe}_{ij}) }, 
\end{equation}
where the estimated posterior edge inclusion probability $\hat{pe}_{ij}$ is the proportion of posterior samples with graph entry $g_{ij}=1$, for $1 \leq i < j \leq p$. Note that the threshold $10^{1/2}$ corresponds to $\hat{pe}_{ij}\approx 0.76$.
We assume equally likely models $p(\hat{g}_{ij}=1)=p(\hat{g}_{ij}=0)=1/2$.
Alternatively, one may want to favor different criteria for the threshold in the Bayes factor. 
Furthermore, the entries of the posterior precision estimate $\hat{\KK}=\{\hat{k}_{ij}\}$ are computed as the mean of the precision samples $k_{ij}$ for all MCMC iterations whose corresponding graph edge $g_{ij}$ is consistent with the marginal posterior graph $\hat{g}_{ij}$. 
We keep all MCMC samples, after warm-up, when computing the posterior mean correlation estimate $\hat{r}_{ij}$. 

\subsection{Implementation}\label{implementation}
We have implemented the proposed model within BEAST \citep{suchard2018}.

\section{Simulation Study}
\label{sec Simulation Study}
In order to understand the behavior of our graphical model (GPTEM) we conduct a simulation study consisting of two scenarios, $Sim \ 1$ and $Sim \ 2$, each combining different graph structures $\GG_0$ and precision matrices $\KK_0$. We also compare these results to the ones obtained from the full model, in which the association structure is not explicitly modeled through a graph. 
We use $N=50$ and $p=5$ for $Sim\ 1$, and $N=100$ and $p=10$ for $Sim\ 2$. The diffusion precision $\KK_0$ used to generate the continuous trait observations in each simulation scenario and the corresponding diffusion correlation $\RR_0$ are provided in Supplementary Material A2 \citep{pinheiro2022}. The true graph structure $\GG_0$ can be directly recovered from the respective $\KK_0$.

For each scenario, we simulate $RE=1000$ Monte Carlo data sets and run MCMC chains to fit both graphical and full models. For each Monte Carlo replication, we generate a random tree $\FF$ of size $N$ and simulate the latent variables $\XX_h$, $h=1,\dots,2N-1$, traversing the tree from root $\XX_{2N-1}$ to tips, using Equation (\ref{eq: X_h}) with diffusion precision $\KK_0$. We set $\YY_k=\XX_k$ for the tips $k=1,\dots,N$. As non-informative priors, we take a uniform distribution over the graph space on $\GG$, and a G-Wishart $\mathcal{W}_G(3, \bf{I}_p)$ on $\KK|\GG$ for the graphical model, whereas we assume a Wishart $\mathcal{W}_p(2+p, \bf{I}_p)$, with rate parametrization, on $\KK$ for the full model, where $\bf{I}_p$ denotes the identity matrix of dimension $p$. The Wishart and G-Wishart hyperparameters are equivalent since $\nu=\delta+p-1$, where $\nu$ and $\delta$ are the degrees of freendom of Wishart and G-Wishart distributions, respectively. 
We approximate the posterior distribution of the graph structure $\GG$ for the graphical model and the posterior distribution of the diffusion precision $\KK$ and diffusion correlation $\RR$ for both models. Simulations are tailored to reach an effective sample size $\text{ESS} \approx 500$ after warm-up. For each Monte Carlo replication $re=1,\dots,RE$, we estimate the posterior graph $\hat{\GG}^{(re)}$ using the Bayes factor criteria and threshold in Eq. (\ref{eq: BF threshold}).

\begin{figure}[htp]
    \centering
    \makebox[\linewidth][c]{
    \subfloat[\label{fig: MC graph}(a) Monte Carlo graphs $\hat{\GG}^{MC}$ and true association structure $\GG_0$.]{\includegraphics[width=.95\textwidth]{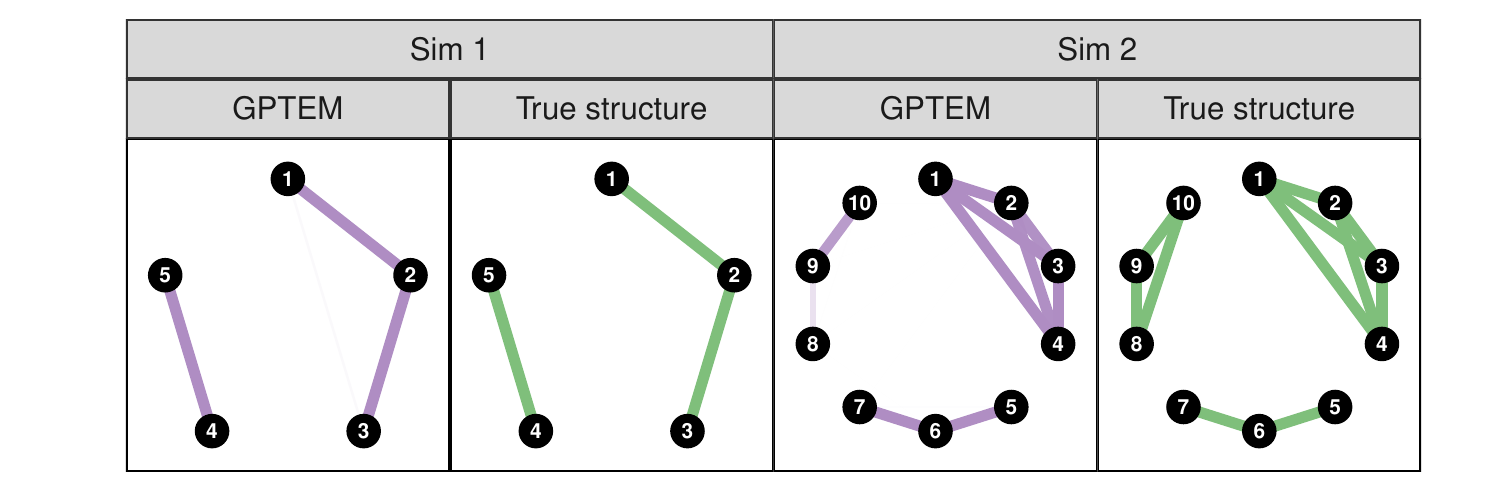}}} \\
    \makebox[\linewidth][c]{
    \subfloat[\label{fig: MC pe}(b) Posterior edge inclusion probabilities $\hat{\bm{pe}}$ across $\RR_0$]{\includegraphics[width=.95\textwidth]{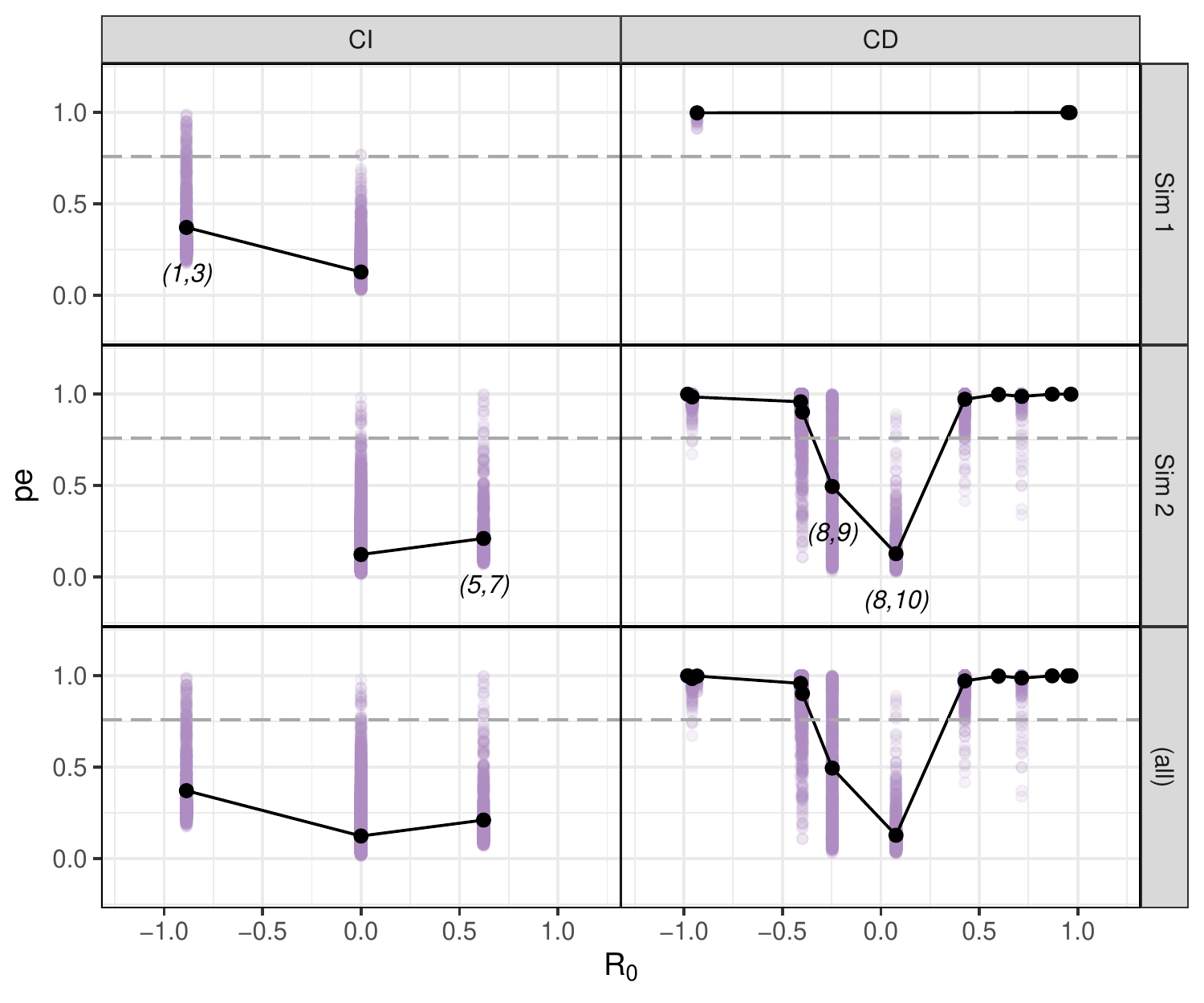}}}
     \caption{\small 
     (\ref{fig: MC graph}) Monte Carlo posterior graph $\hat{\GG}^{MC}$ and true graph $\GG_0$ in both simulation scenarios. Edge thickness and transparency represent the proportion of Monte Carlo replicates that include edge $(i,j)$ in the posterior graph. (\ref{fig: MC pe}) Posterior edge inclusion probabilities of each Monte Carlo replication across $\RR_0$. Colored points depict $\hat{pe}^{(re)}_{ij}$ the proportion of MCMC samples in which edge $\hat{g}_{ij}=1$ for each 1000 Monte Carlo replicate. The black points represent the $\hat{pe}^{MC}_{ij}$ of each pairwise correlation. Edges $(1,3)$ from $Sim \ 1$ and $(5,7)$, $(8,9)$ and $(8,10)$ from $Sim \ 2$ are indicated to illustrate important features of graph estimation.}
     \label{graph panel}
\end{figure}

Figure \ref{graph panel} presents the Monte Carlo posterior graph $\hat{\GG}^{MC} =\{ \hat{g}^{MC}_{ij} \}$, calculated as $\hat{\GG}^{MC} = (RE)^{-1}\sum_{re = 1}^{RE} \hat{\GG}^{(re)}$ and the respective true graph structure for both simulation scenarios.
Note that the Monte Carlo graph is a summary measure for the graph estimates in each Monte Carlo replication, not a direct estimate itself. Therefore, in Figure \ref{fig: MC graph}, edge thickness and transparency represent $\hat{g}^{MC}_{ij}$, i.e.\ the proportion of Monte Carlo replicates that include edge $(i,j)$ in the posterior graph. 
In $Sim\ 1$, $\hat{\GG}^{MC}$ perfectly matches the true graph structure, which means that Monte Carlo replicates correctly estimate the association structure. 
In $Sim\ 2$, however, edges $(8,9)$ and $(8,10)$ are not always included in the posterior graphs replicates (Figure \ref{fig: MC graph}).

To help elucidate the inconsistencies in graph estimation, we show, in Figure \ref{fig: MC pe}, the posterior edge inclusion probabilities for each Monte Carlo replication $\hat{pe}_{ij}^{(re)}$ (colored dots), as well as their means $\hat{pe}^{MC}_{ij}$ (black dots). We additionally stratify the edges by conditional (in)dependence type --- as conditionally independent (CI), when the true graph ${g_{0}}_{ij}=0$, or conditionally dependent (CD), when ${g_{0}}_{ij}=1$ ---, and plot them against the true correlation $\RR_0$. The last grid combines $\hat{\bm{pe}}$ from both simulations to facilitate the overall visualization, but should be interpreted with care, since sample sizes and trait dimensions are different between simulations. 

We single out a few edges in Figure \ref{fig: MC pe} in order to highlight a couple of model features. 
Note that edge $(1,3)$ in $Sim\ 1$ has mean posterior edge inclusion probability $\hat{pe}^{MC}_{13}=0.372$ (black dot for $(1,3)$), but mean posterior graph entry $\hat{g}^{MC}_{13}=0.051$, which indicates that only 5.1\% of the Monte Carlo replications display posterior edge inclusion probability greater than or equal to 0.76 (see the proportion of colored dots above the dashed lines in Figure \ref{fig: MC pe}). This result suggests that the absence of edges for CI variables can be correctly estimated even when the respective true correlation is very strong (e.g. ${r_0}_{13}=-0.89$).
For CD variables, on the other hand, one can see that posterior edge inclusion probabilities are high for strong true correlations, but they start decreasing as the true correlations approach zero (Figure \ref{fig: MC pe}). 
This is the case of edges $(8,9)$ and $(8,10)$ in $Sim \ 2$ that display weak correlations, namely, $-0.25$ and $0.08$, respectively.

\begin{figure}[hb]
\centering
  \includegraphics[width=0.9\linewidth]{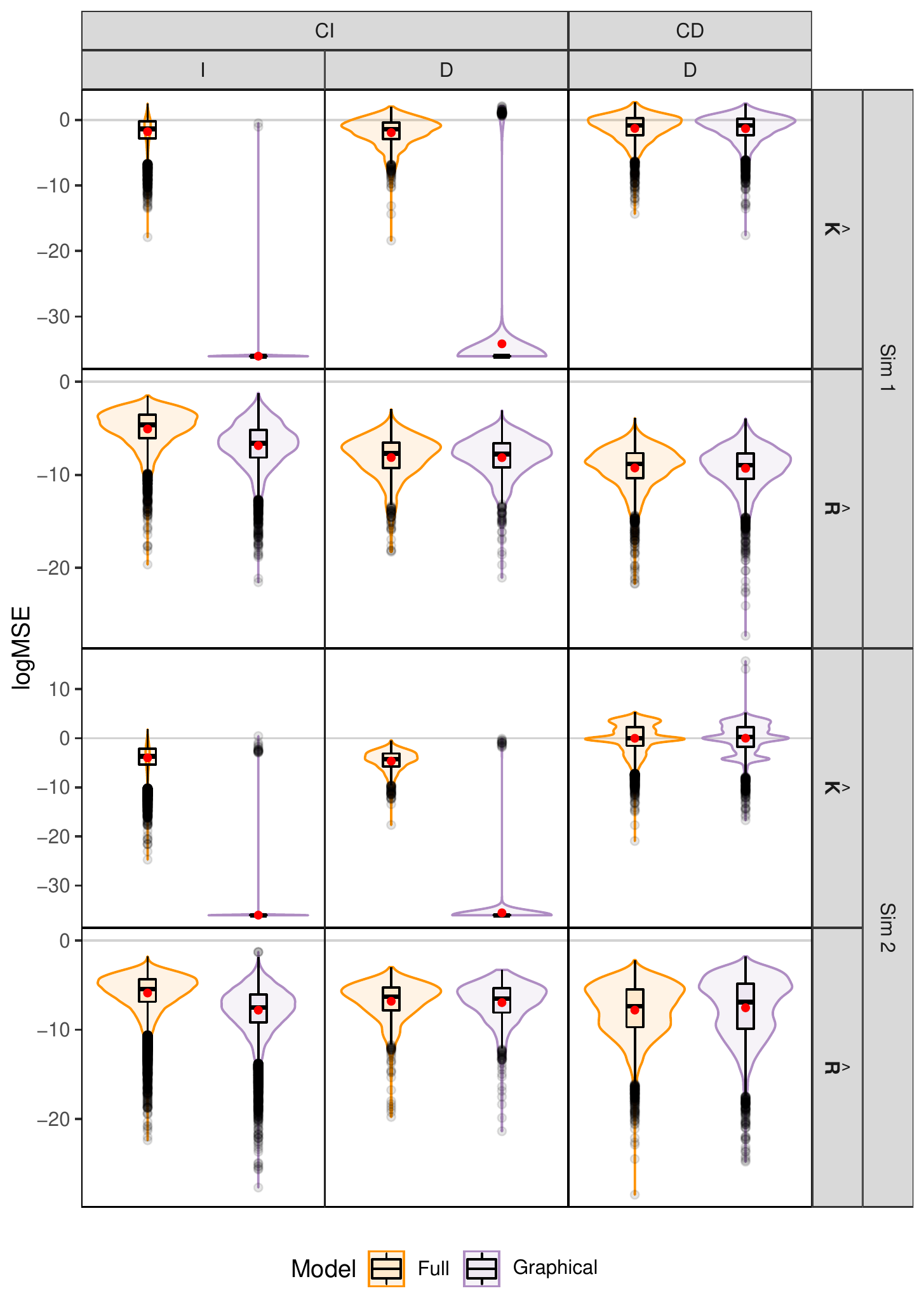}
  \caption{\small Log mean squared-error of posterior diffusion precision $\hat{\KK}^{(re)}$ and posterior diffusion correlation estimates $\hat{\RR}^{(re)}$ over 1000 simulated replicates based on $Sim\ 1$ and $Sim\ 2$. Red dots represent the Monte Carlo mean $log\text{MSE}$. CI = conditionally independent; CD = conditionally dependent; I = independent; D = dependent.}
  \label{MSE}
\end{figure}

Figure \ref{MSE} presents the log mean squared-error $log\text{MSE}$ for precision $\hat{\KK}^{(re)}$ and correlation $\hat{\RR}^{(re)}$ estimates in both models. For this analysis we classify each parameter entry according to its conditional (in)dependence type (CI or CD) and, additionally, by its dependence type --- as independent (I), if the true correlation is zero ${r_0}_{ij}=0$, or dependent (D), if ${r_0}_{ij} \neq 0$. Then, the categories are: conditionally independent and independent (CI-I), conditionally independent and dependent (CI-D), and conditionally dependent and dependent (CD-D). 

As expected, in both simulation scenarios, the mean precision $log\text{MSEs}$ for CI variables (CI-D and CI-I) are significantly lower in the graphical model, specially for the independent case (CI-I). This is an important result since CI variables are the main target for sparsity in GGM. 
Note that, for CD variables, the $log\text{MSEs}$ are equivalent between the models, which is consistent with the fact that Wishart and G-Wishart sampling processes are the same for CD variables \citep[for details, see][]{atay2005}. 
The mean correlation $log\text{MSE}$ for CI-I variables are also lower in graphical models. This result suggests that the better estimates for precision matrices in the graphical model lead to lower $log\text{MSEs}$ for correlations in this class of variables. Additionally, note that correlations between dependent variables (CI-D and CD-D) display similar log mean squared-errors regardless their conditional (in)dependence status, in both simulations.

Next, we compare model selection criteria. Since the full model relies on a post-processing procedure over the posterior distribution of $\RR$ to perform correlation selection, and the graphical model directly estimate $\hat{\GG}$, we compute accuracy, sensitivity, specificity, precision, and F1-Score for the respective model target parameter in order to establish a fair comparison between models. These metrics depict the overall performance of parameter estimation. However, since they are derived from confusion matrices, which are proper to evaluate binary classification systems, for the full model, we define the target parameter in each Monte Carlo iteration as the binary matrix $\hat{\RR}^{H(re)}$, obtained by classifying each marginal correlation into zero or non-zero using a specific HPD criteria. We define
\begin{equation*}
    \hat{r}_{ij}^{H(re)}:=
        \begin{cases}
            1, & \text{if } 0 \notin \text{HPD}_{\gamma\%}(r_{ij}^{(re)}) \\
            0, & \text{otherwise},
        \end{cases}
\end{equation*}
where $\gamma$ indicates the chosen percentage for HPD criteria. 

Table \ref{tab confusion} reports the comparisons of our graphical model with the full model applying HPD$_{90\%}$ and HPD$_{95\%}$ criteria.
Our method performs well overall as its specificity, F1-score, precision and accuracy are higher than the full model in both simulations. Sensitivity is one for all models and criteria in $Sim\ 1$, which means that all the true edges are correctly identified in the graphical model as well as all the true non-zero correlations in the full model for both HPD criteria. In $Sim\ 2$, sensitivity is higher for the full model with HPD$_{90\%}$ ($0.91\pm 0.054$).

\begin{table}[b]
  \caption{\small Summary of performance measures for decision criteria in graphical and full models. In the full model, the decision is to classify the correlations as zero or non-zero according to a chosen HPD; in the graphical model, it is whether the edges should be included in the posterior graph or not. The table presents the mean (and standard deviation) for sensitivity, specificity, precision, F1-score, and accuracy of posterior graph estimates $\hat{\GG}^{(re)}$ (graphical model) and binary correlation estimates $\hat{\RR}^{H(re)}$ (full model). 
  The best model for each statistic is boldfaced.}
  \smallskip
  \resizebox{\textwidth}{!}{
  \small
  \begin{tabular}{lllcccccccccc}
    \hline
    \textbf{Simulation} & 
    \textbf{Parameter} & 
    \textbf{Criteria} & 
    \textbf{Sensitivity} & 
    \textbf{Specificity} &
    \textbf{Precision} &
    \textbf{F1-Score} & 
    \textbf{Accuracy} \\
    \hline
    \multirow[t]{3}{*}{${Sim\ 1}$} & $\hat{\GG}$   & GGM   & \textbf{1.00}  (0.000) & \textbf{0.99} (0.032) & \textbf{0.99} (0.056) & \textbf{0.97} (0.059) & \textbf{0.98} (0.042) \\
    & $\hat{\RR}^{H}$   & HPD$_{90\%}$ & \textbf{1.00}  (0.000) &  0.72  (0.266) & 0.66 (0.155) & 0.79 (0.132) & 0.81  (0.186) \\
    & $\hat{\RR}^{H}$   & HPD$_{95\%}$ & \textbf{1.00}  (0.000) &  0.78  (0.210) & 0.70 (0.127) & 0.81 (0.107) & 0.84  (0.147) \\
    \hline
    \multirow[t]{3}{*}{${Sim\ 2}$} & $\hat{\GG}$   & GGM   & 0.82  (0.055) & \textbf{1.00} (0.005) & \textbf{1.00} (0.018) & \textbf{0.90} (0.034) & \textbf{0.96} (0.014) \\
    & $\hat{\RR}^{H}$   & HPD$_{90\%}$ & \textbf{0.91}  (0.054) & 0.85 (0.093) & 0.70 (0.132) & 0.78  (0.092) & 0.87  (0.072) \\
    & $\hat{\RR}^{H}$   & HPD$_{95\%}$ & 0.90   (0.055) & 0.91  (0.072)& 0.78 (0.120) & 0.83  (0.079) & 0.91  (0.056) \\
    \hline
\end{tabular}}
  \label{tab confusion}
\end{table}
Since $\RR_0$ in $Sim \ 2$ (see $\RR_0$ (A2.4) in Supplementary Material A \citep{pinheiro2022}) displays some weak correlations that tend to be shrunk by HPD correlation selection in the full model --- or equivalently, whose corresponding graph entries are pushed towards zero in the graphical model  ---, the apparent best performance in sensitivity of HPD$_{90\%}$ criteria is actually a consequence of its less strict interval that favors non-zero correlation classifications, even for weak true correlations, at the cost of less specificity ($0.85\pm 0.093$). 

Overall, the graphical model is better at identifying CI and CD variables than the HPD criteria are at discriminating between correlated and uncorrelated variables in the full model. 
This is particularly the case of the precision metric, that depicts the proportion of true positives in the confusion matrix among all predicted as true (including the false positives), and F1-Score, which summarizes the performances by balancing sensitivity and precision. One can see that, across simulations, the precision metric for $\hat{\GG}^{(re)}$ is at least 0.99 in the graphical model, but lies under 0.78 for $\hat{\RR}^{H(re)}$ in full models, indicating that the number of false positives is relatively high for HPD procedures.

We also compute the accuracy of $\hat{\GG}$ and $\hat{\RR}^{H}$ directly on each parameter entry, stratifying the variables by the combination of conditional (in)dependence type (CI or CD) and dependence type (I or D). This statistic provides better insight about how the conditional (in)dependence structure and correlation type can affect model selection.

In both simulation scenarios, the overall accuracy for each entry is higher for $\hat{\GG}$ in the graphical model than it is for the analogous $\hat{\RR}^{H}$ in the full model (Table \ref{tab: accuracy}). The graphical model also displays higher average accuracy for CI-I variables, which is one of the substantial gains of including the graph estimation in the MBD model. Note that the difference in accuracy between the models is highlighted by the fact that standard deviations in CI-I variables are relatively small in both simulation scenarios. 
On the other hand, for CI-D variables, despite being high in the graphical model, the average accuracy is larger for the full model in both simulations. A possible reason for that is the underlying strong correlation of CI-D variables in our simulation settings (see edges $(1,3)$ in $Sim\ 1$, and $(5,7)$ in $Sim\ 2$, Figure \ref{fig: MC pe}) that might negatively affect graphical model performance.

\begin{table}[b]
  \centering
  \caption{\small Summary of Monte Carlo accuracy measures on each parameter entries in two simulations scenarios for graphical and full models. The table presents the average (and standard deviation) accuracy of posterior graph $\hat{\GG}^{(re)}$ edge estimations in the graphical model, and posterior binary correlation $\hat{\RR}^{H(re)}$ entries in the full model. The number of pairwise entries classified in each category according to the conditional (in)dependency (CD or CI) and dependency types (D or I) is indicated by $n$. The best models are boldfaced.}
    \begin{tabular}{lllllclccl}
    \hline
    \multicolumn{1}{c}{\multirow{2}{*}{ \textbf{Model}} } &
    \multicolumn{1}{c}{\multirow{2}{*}{ \textbf{Parameter}} } &
    \multicolumn{1}{c}{\multirow{2}{*}{ \textbf{Criteria} } } & \multicolumn{1}{c}{\multirow{2}{*}{ $\GG_0$} } & \multicolumn{1}{c}{\multirow{2}{*}{$\RR_0$}} & 
    \multicolumn{2}{c}{$Sim\ 1$} & & \multicolumn{2}{c}{$Sim\ 2$}   \\
    \cline{6-7}\cline{9-10} & & & & & 
    \multicolumn{1}{c}{n} & \multicolumn{1}{c}{\textbf{Accuracy (SD)}} &       & \multicolumn{1}{c}{n} & \multicolumn{1}{c}{\textbf{Accuracy (SD)}} \\
    \hline
    Graphical & $\hat{\GG}$    & GGM             & CD          & D         & 3     & \textbf{1.00} (0.000)     &   & 11    & 0.82 (0.348) \\
    Full & $\hat{\RR}^{H}$  & HPD$_{90\%}$    & CD          & D         & 3     & \textbf{1.00} (0.000)     &   & 11    & \textbf{0.91} (0.236) \\
   
    Full & $\hat{\RR}^{H}$  & HPD$_{95\%}$    & CD          & D         & 3     & \textbf{1.00} (0.000)     &   & 11    & 0.90 (0.264) \\ 
    \hline
    Graphical & $\hat{\GG}$    & GGM             & CI          & D         & 1     & 0.95                     &   & 1     & 0.99  \\
    Full & $\hat{\RR}^{H}$  & HPD$_{90\%}$    & CI          & D         & 1     & \textbf{1.00}            &   & 1     & \textbf{1.00} \\
    Full & $\hat{\RR}^{H}$  & HPD$_{95\%}$    & CI          & D         & 1     & \textbf{1.00}            &   & 1     & \textbf{1.00} \\
    \hline
    Graphical & $\hat{\GG}$    & GGM             & CI          & I         & 6     & \textbf{1.00} (0.001)     &   & 33    & \textbf{1.00} (0.001) \\
    Full & $\hat{\RR}^{H}$  & HPD$_{90\%}$    & CI          & I         & 6     & 0.84 (0.007)              &   & 33    & 0.88 (0.010) \\
    Full & $\hat{\RR}^{H}$  & HPD$_{95\%}$    & CI          & I         & 6     & 0.90 (0.007)              &   & 33    & 0.94 (0.007) \\
    \hline
    Graphical & $\hat{\GG}$    & GGM             & Overall     & Overall   & 10    & \textbf{1.00} (0.016)     &   & 45    & \textbf{0.96} (0.183) \\
    Full & $\hat{\RR}^{H}$  & HPD$_{90\%}$    & Overall     & Overall   & 10    & 0.91 (0.082)              &   & 45    & 0.89 (0.115) \\
    Full & $\hat{\RR}^{H}$  & HPD$_{95\%}$    & Overall     & Overall   & 10    & 0.94 (0.050)              &   & 45    & 0.93 (0.128) \\
    \hline
    \end{tabular}
  \label{tab: accuracy}%
\end{table}%

In addition, both models and criteria performed well in terms of pairwise accuracy for CD-D variables. Despite the better accuracy for $\hat{\RR}^{H}$ with the HPD$_{90\%}$ criteria in the full model in $Sim\ 2$, all three statistics are similar, specially when considering their high standard deviations. Finally, in a general perspective, these accuracy results corroborate the ones obtained for the $log$MSE of $\hat{\KK}$ where better estimates for the graphical model are associated with CI-I and similar results between models are obtained for dependent variables (CI-D and CD-D).

\section{Applications}
\label{sec Application}
We apply our method to two data sets to showcase the benefits of including graph estimation in the MBD model and to demonstrate how the association structure represented by the graph can lead to richer discussions when compared to traditional trait evolution models. For each application, we run both graphical and full models in equivalent setups. We choose HPD$_{95\%}$ as the criteria for full model HPD correlation selection.

\subsection{Darwin's Finches}\label{subsec Darwins}
Evolution of Darwin’s finches (Fringillidae, Passeriformes) is a classical example of adaptive radiation under natural selection. There are thirteen species in the Galápagos archipelago and another one on Cocos island. 
The wide variation in beak morphology is associated with the exploitation of a variety of ecological niches, because it allows finches to access particular types of food --- including seeds, insects, and cactus flowers \citep{abzhanov2004, abzhanov2006} ---, and likely played a role in the diversification of avian species \citep{mallarino2011}.
To assess the phenotypic correlations and the association structure between morphometric variables, we use the data set of 13 species of Darwin’s finches from \citet{drummond2012}. The data consist of a 2,065-bp partial nucleotide alignment of the mitochondrial control region and cytochrome b genes and five continuously measured phenotypic traits: culmen length (\textit{CulmenL}), beak depth (\textit{BeakD}), gonys width (\textit{GonysW}), wing length (\textit{WingL}), and tarsus length (\textit{TarsusL}).
We estimate the posterior graph $\hat{\GG}$ for the graphical model and the diffusion correlation $\hat{\RR}$ in both graphical and full models in order to illustrate the gain of explicitly accounting for the association structure in our inference framework.

\begin{figure}[htp]
    \centering
    \makebox[\linewidth][c]{
    \subfloat[\label{darwin G draw}(a) Graphical model $\hat{\GG}$]{\includegraphics[width=0.72\textwidth]{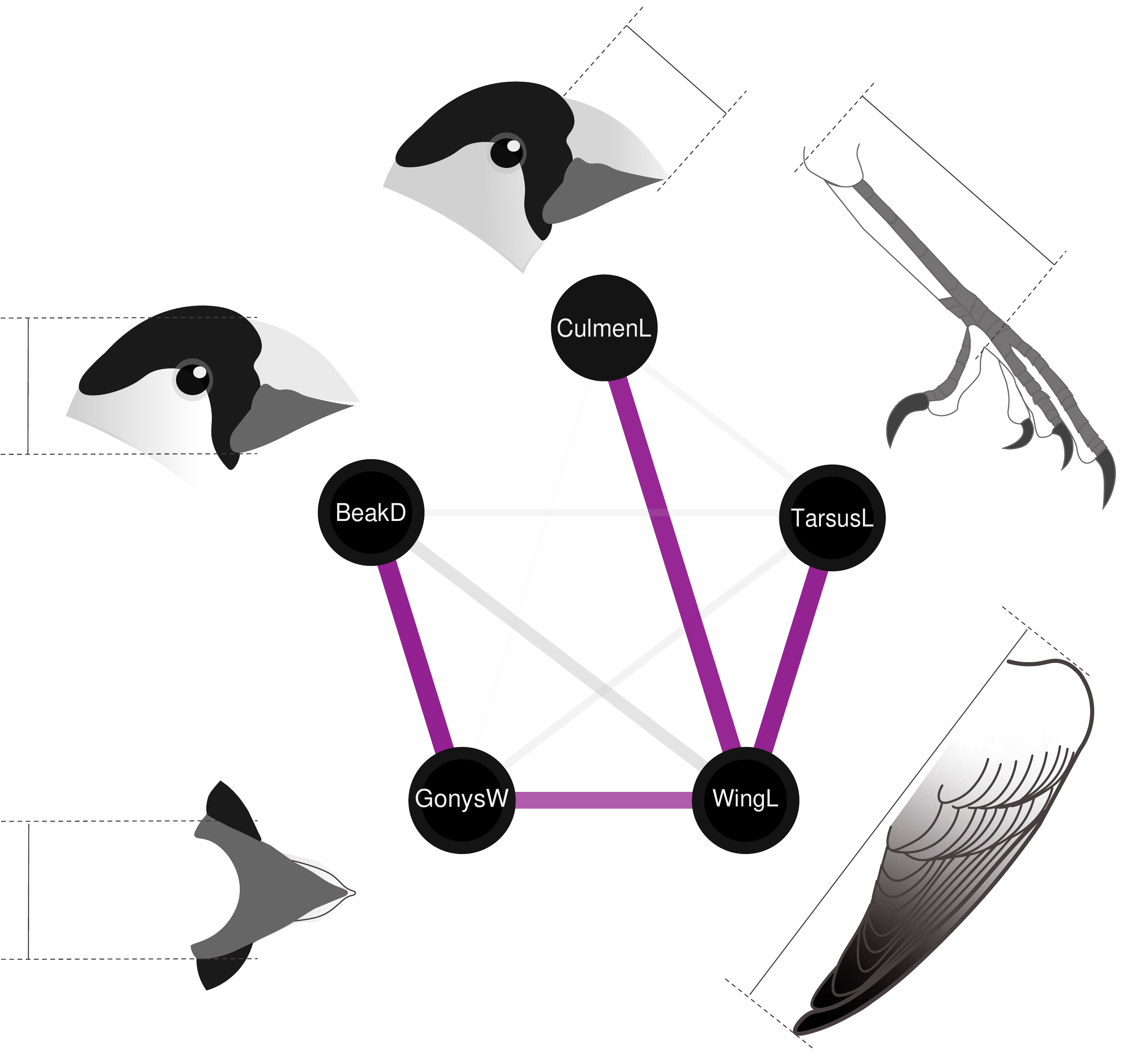}}}
    \makebox[\linewidth][c]{
    \subfloat[\label{darwin RG}(b) Graphical model $\hat{\RR}$ and $\bm{pe}$]{\includegraphics[width=0.5\textwidth]{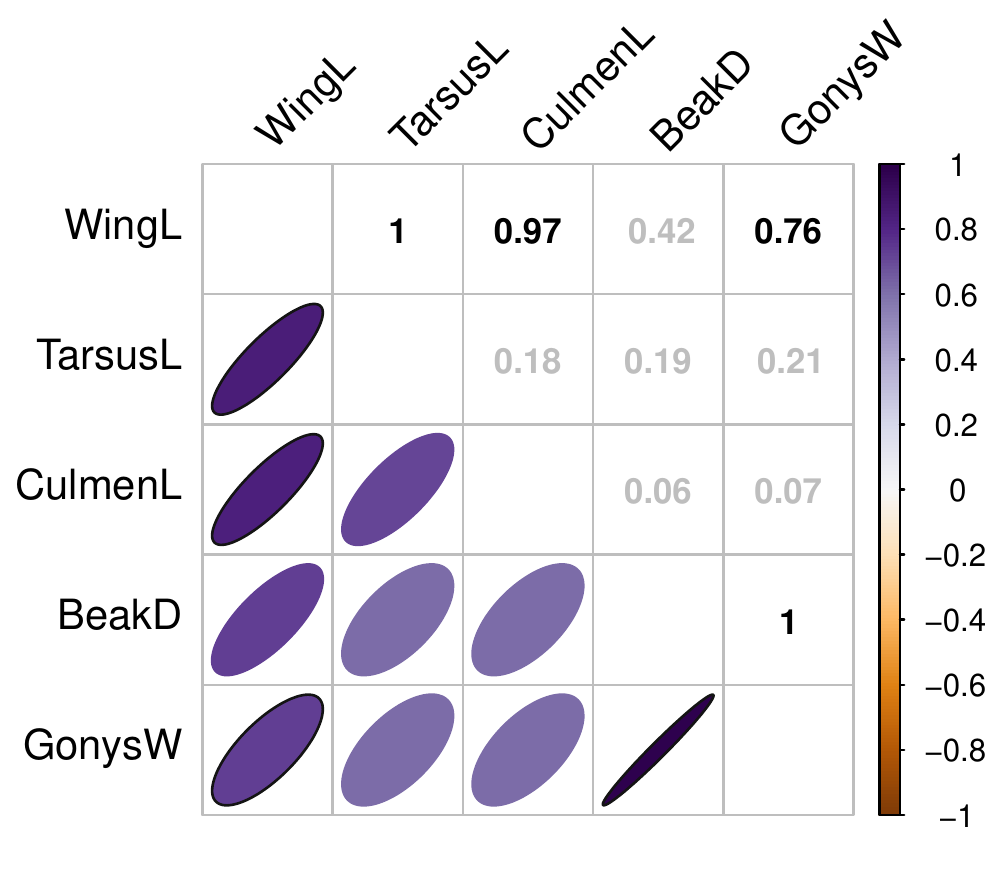}}
    \subfloat[\label{darwin R}(c) Full model $\hat{\RR}$ and $\bm{ps}$]{\includegraphics[width=0.5\textwidth]{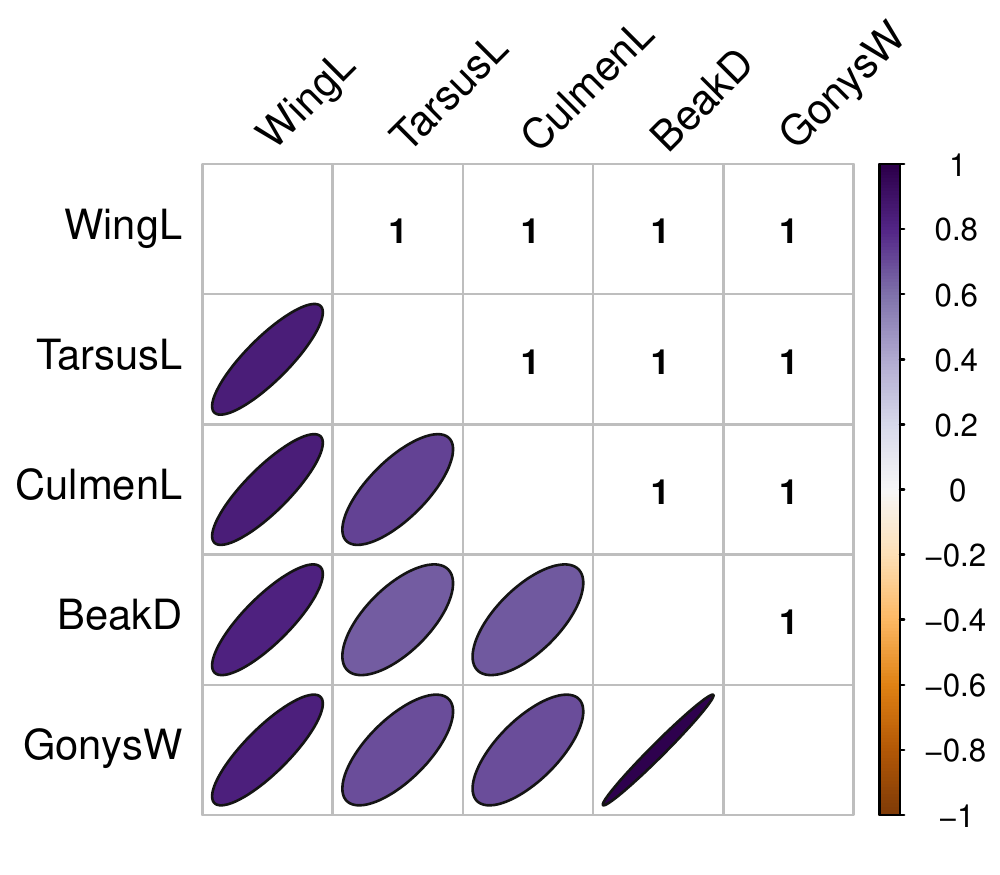}}}
     \caption{Association structure and correlation between Darwin's Finches morphometric traits in graphical and full models. (\ref{darwin G draw}) Posterior Graph. Graph edge thickness and transparency represent the posterior edge inclusion probability $\bm{pe}$ in the graphical model. (\ref{darwin RG}) (\ref{darwin R}) Correlograms for graphical and full models.
     The ellipses below correlogram diagonal summarize the posterior mean correlations $\hat{r}_{ij}$ between each pair of traits. In the graphical model, the numbers above the diagonal report  the posterior edge inclusion probability $\bm{pe}$ (Figure \ref{darwin RG}), while in the full model they report  the posterior probability that the correlation is of the same sign as its mean $\bm{ps}$, as an alternative visualization of HPD (Figure \ref{darwin R}). Bold numbers above diagonal indicate included edges in the posterior graph (graphical model) or significant correlations using HPD$_{95\%}$ (full model).}
     \label{darwin}
\end{figure}

Figure \ref{darwin G draw} displays the estimated posterior graph $\hat{\GG}$ for the association structure between Darwin's finches trait measurements, whereas Figures \ref{darwin RG}, and \ref{darwin R} present the evolutionary correlation between those traits in the graphical and full models, respectively. The numbers above diagonal in each correlogram represent the posterior edge inclusion probabilities $\bm{pe}$ in the graphical model, and the posterior probability that correlations are of the same sign of its mean $\bm{ps}$, which is a proxy for the percentage whose HPD interval would not contain zero.

The correlograms of both graphical and full models show similar estimates, although pairwise correlations are slightly stronger in the full model (Figure \ref{darwin RG} and \ref{darwin R}). 
From the full model perspective, all ten pairwise correlations are significant, since $\bm{ps}=1$ for each trait combination (Figure \ref{darwin R}). Correlations are all positive and relatively strong, varying between 0.65 and 0.99.
Interestingly, note that stronger correlations correspond to the pairwise variables that share an edge in the graphical model graph.
Notice also that the association structure represented by the estimated graph strongly enhances our ability to explore the intricated correlations presented in the correlograms.

From Figure \ref{darwin G draw} one can see that the wing length is directly associated with tarsus length ($\hat{r}_{\textit{WingL},\textit{TarsusL}}=0.84$, $pe=0.99$), culmen length ($\hat{r}_{\textit{WingL},\textit{CulmenL}}=0.83$, $pe=0.97$), and gonys width ($\hat{r}_{\textit{WingL},\textit{GonysW}}=0.74$, $pe=0.76$). However, conditioning on the wing length and the remaining variables, culmen length, tarsus length and gonys width are pairwise independent, which suggest that there is no evidence for direct interaction between these traits during their evolution among analysed finch species. Additionally, the beak depth is directly and exclusively associated with gonys width ($\hat{r}_{\textit{BeakD},\textit{GonysW}}=0.98$, $pe=1$), highlighting their decisive conditional dependency given the other variables in the graphical model.
The conditional independence found between culmen length and beak depth ($pe=0.06$) and culmen length and gonys width ($pe=0.07$) couple with the findings of a sequence of studies on the identification of a regulatory network governing the morphology of the prenasal cartilage (\textit{pnc}) \citep{abzhanov2004,abzhanov2006} in Darwin's Finches.

\citet{abzhanov2004} found that the expression of the bone morphogenetic protein 4 (\textit{Bmp4}) in the mesenchyme of the upper beaks strongly correlated with deep and broad beak morphology, explaining the linkage in their variation. This \textit{Bmp4} regulatory pathway could explain why beak depth and gonys width share an edge in the posterior graph. However, it is important to consider that those results were obtained for the upper beak only, and our finches data set provide only: i) gonys width measures, which correspond to the lower beak; and ii) beak depth without discriminating between upper and lower parts. Therefore, a more detailed data set is required to build a proper causal connection between the conditional dependency found in the posterior graph and the \textit{Bmp4} regulatory pathway in lower beaks measurements.

Additionally, \citet{abzhanov2006} found that local upregulation of the calmodulin-dependent (\textit{CaM}) pathway is likely to have been a component of the evolution of Darwin’s finch species with elongated beak morphology and provide a mechanistic explanation for the conditional independence of beak evolution between length and width/depth axes. Both \textit{Bmp4} and \textit{CaM} regulate morphogenesis of the prenasal cartilage (\textit{pnc}) in early development, which forms the initial beak skeleton. However, much of the beak diversity in birds depends on variation in the premaxillary bone (\textit{pmx}), that forms later in development and becomes the most prominent functional and structural component of the adult upper beak/jaw. Therefore, although those studies might explain just a fraction of the complex beak measurement variation in Finches, they are definitely congruent with the association structure estimated by GPTEM.

\subsection{Prokaryote Evolution}\label{subsec Prokaryotes}
We revisit the application of \citet{hassler2020} concerning correlation estimates among a set of genotypic and phenotypic prokaryote traits.
This data set collects information for $N=705$ prokaryotes, combining cell diameter (\textit{CellD}), cell length (\textit{CellL}), optimum temperature (\textit{Topt}), and pH measurements (\textit{pH}) from \citet{goberna2016}, as well as data on genome length (\textit{GenL}), coding sequences length (\textit{CDS}), and GC content (\textit{GC}) from the prokaryotes table in NCBI Genome. We use 16S sequences to infer the phylogeny and after fitting graphical and full models, according to model setup described in \citet[Section 7.2]{hassler2020},  we obtain posterior samples for parameters of scientific interest. We discard the first 20\% of the samples as warm-up. From the posterior distribution of $\RR$ and $\GG$ we estimate the posterior graph $\hat{\GG}$, and the posterior correlation $\hat\RR$. 

\begin{figure}[htp]
    \centering
    \makebox[\linewidth][c]{
    \subfloat[\label{prok G draw}(a) Marginal posterior graph $\hat{\GG}$]{\includegraphics[width=\textwidth]{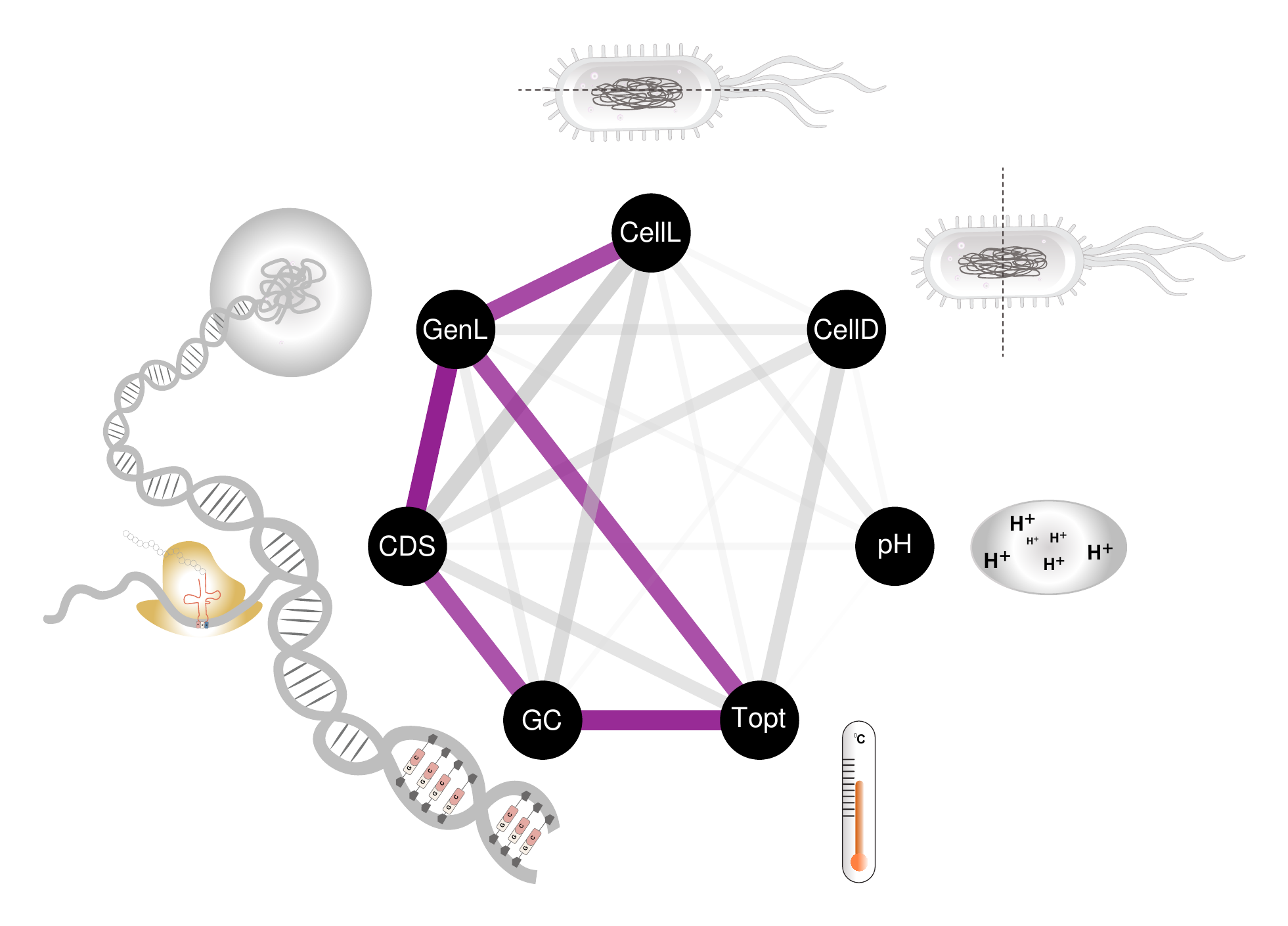}}
    }\\
    \makebox[\linewidth][c]{%
    \subfloat[\label{prok RG}(b) Graphical model $\hat{\RR}$ and $\bm{pe}$]{\includegraphics[width=0.5\textwidth]{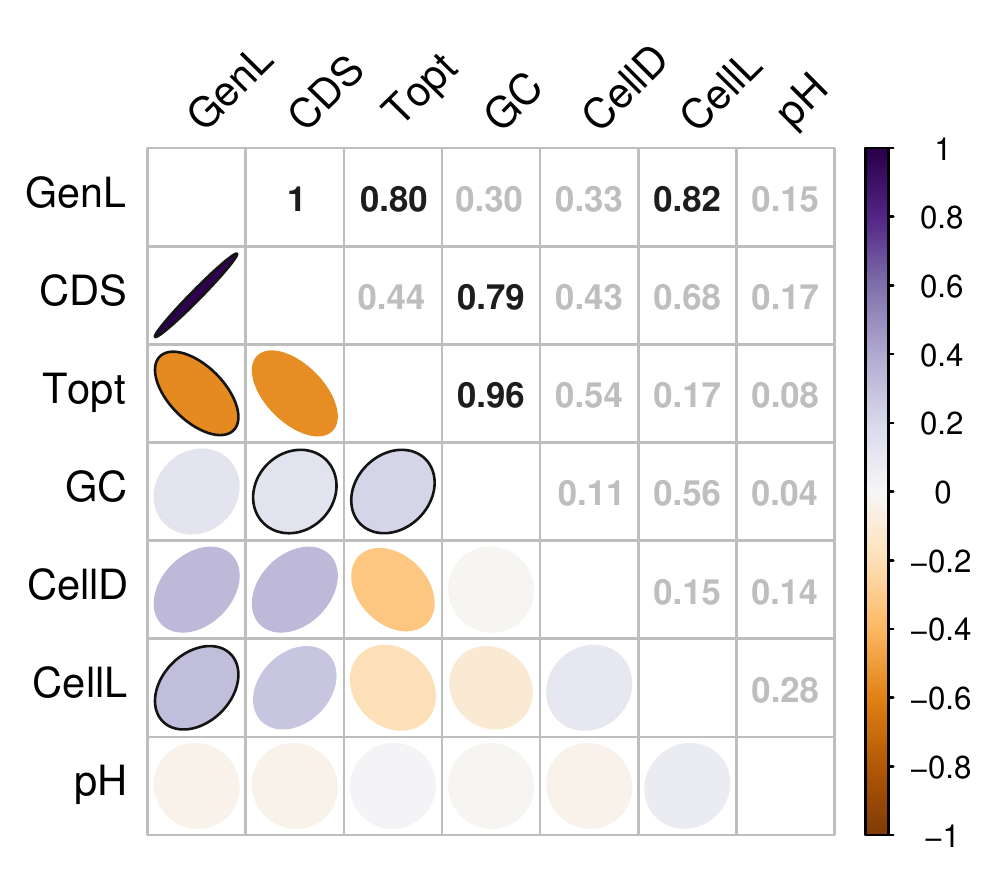}}
    \subfloat[\label{prok R}(c) Full model $\hat{\RR}$ and $\bm{ps}$]{\includegraphics[width=0.5\textwidth]{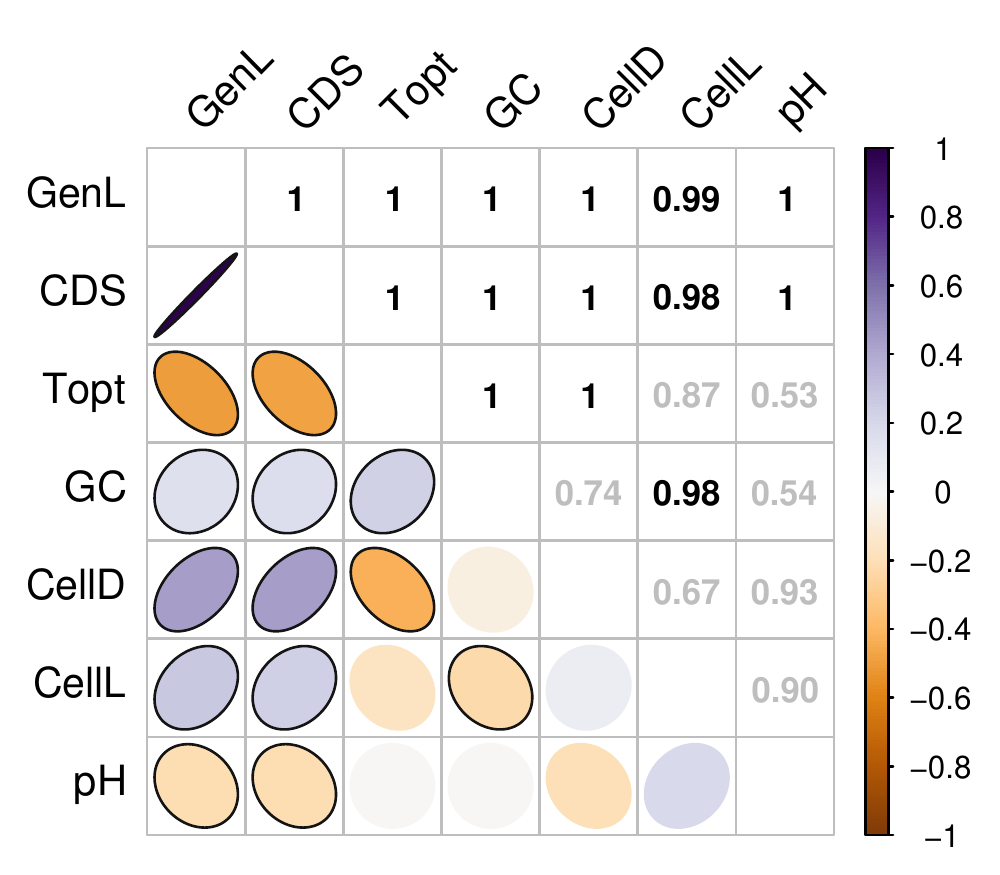}}}
 \caption{ Association structure and correlation among prokaryotic growth properties. See Figure \ref{darwin} for caption. In the graphical model, marked ellipses and numbers correspond to the edges included in the marginal posterior graph $\GG$, while in the full model they highlight the significant correlations according to the HPD$_{95\%}$ criteria.}
\end{figure}\label{prokatyotes}

Figure \ref{prok G draw} displays the estimated posterior graph $\hat{\GG}$ with the associations between trait measurements, and Figure \ref{prok RG} and \ref{prok R} present the posterior correlation matrix $\hat{\RR}$ in both graphical and full models. In the full model, marked ellipses and upper diagonal numbers in bold single out significant correlations using HPD$_{95\%}$ criteria to perform correlation selection, whereas, in the graphical model, they indicate that corresponding edges are included in the posterior graph.
By comparing both correlograms, one can see that correlations are similar for most coefficients. However, we see an important divergence between the models in the correlation of pH with both genome length and coding sequence length. While the full model identifies those correlations as significant ($\hat{r}_{\textit{pH},\textit{GenL}}=-0.20$, HPD$_{95\%}=\left[ -0.34, -0.05 \right]$ and $\hat{r}_{\textit{pH},\textit{CDS}}=-0.20$, HPD$_{95\%}=\left[ -0.35, -0.06 \right]$), the graphical model shrinks them towards zero ($\hat{r}_{\textit{pH},\textit{GenL}}\approx \hat{r}_{\textit{pH},\textit{CDS}}=-0.04$). The posterior graph shows that \textit{pH} is conditionally independent to all other traits in the graphical model (Figure \ref{prok G draw}), suggesting that pH may have not coupled with any of them during prokaryote evolution.  

The graph structure suggests that the optimum temperature might play an important role as a selective pressure in prokaryote evolution. This is consistent with two well known hypothesis namely genome streamlining \citep{sela2016} and thermal adaptation \citep{bernardi1986}, which discuss the association of optimum temperature with genome size and GC content, respectively, on the evolution of prokaryotes.

One can see that higher optimal temperatures are directly associated with smaller genome length ($\hat{r}_{\textit{Topt},\textit{GenL}}=-0.56$, $pe=0.80$), and smaller genomes are also directly associated with reduced coding sequences (\textit{CDS}), as informed by their conditional strong dependence with high edge inclusion probability in the graph, and the extreme positive correlation between those traits ($\hat{r}_{\textit{GenL},\textit{CDS}}=0.99$, $pe=1$). 
Although \textit{CDS} and \textit{Topt} display a relatively strong negative correlation ($\hat{r}_{\textit{Topt},\textit{CDS}}=-0.56$), the estimated graph indicates that they might actually be conditionally independent given the genome length and the remaining traits in the model ($pe_{\textit{Topt},\textit{CDS}}=0.44$). In addition, \textit{GenL} is also directly associated and positively correlated with the cell length ($\hat{r}_{\textit{GenL},\textit{CellL}}=0.32$, $pe=0.82$). This result suggest that optimal temperature can indirectly affect \textit{CellL} and \textit{CDS} through its direct effect on genome length. Those results corroborate with the genome streamlining hypothesis which states that certain prokaryotic genomes tend to be small in size, due to selection against the retention of non-coding DNA, and in favor of faster replication rates \citep{sela2016}.

Moreover, we find optimal temperature to be also directly associated with the GC content in prokaryotic genomes displaying a positive correlation ($\hat{r}_{\textit{Topt},\textit{GC}}=0.21$, $pe=0.96$). This result points to the polemic thermal adaptation hypothesis which posits that higher GC content is involved in adaptation to high temperatures because it may offer thermostability to genetic material \citep{bernardi1986}.
Additionally, in a recent study, \citet{hu2022} showed positive correlations between optimal growth temperature (\textit{Topt}) and GC content (\textit{GC}) both in bacterial and archaeal structural RNA genes and in bacterial whole genome sequences, chromosomal sequences, plasmid sequences, core genes, and accessory genes, providing additional support for the thermal adaptation hypothesis.

\section{Discussion}
\label{sec: Discussion}
We present a Bayesian inference framework to perform model-based covariance selection, using Gaussian graphical models, in order to learn the association structure between continuous traits while jointly inferring trait evolutionary correlations and the phylogenetic tree. 
By doing so, we introduce a new parameter of scientific interest, the diffusion graph $\GG$, that complements the information provided by the trait correlations estimated in PTE models.

Our approach significantly improves upon traditional trait evolution models in terms of modeling and inference. 
As shown in the simulation study, our model provides better estimates for the diffusion precision and diffusion correlation, specially for independent variables ---  which are the main target for sparsity ---, while displaying similar precision and correlation $log\text{MSEs}$ for dependent variables (CI-D and CD-D). Additionally, GPTEM can accurately identify the association structure between traits. The statistical performance for graph estimation in the graphical model is better than the HPD correlation selection in the full model, which highlights the advantages of a model-based approach for covariance selection. 

When applying the full model we can identify significantly correlated traits, discuss the strength of these correlations, and use it to guide the search for potential biological explanations --- in a possibly dense correlogram.
However, because in many cases the correlations might capture indirect effects of the association structure, simply analyzing the correlograms might leave out nuances in trait relationships.
On the other hand, the graphical model additionally informs about the association structure between traits, which is encoded in the estimated diffusion graph. 
The association structure can help us refine the search for potential mechanisms to explain the conditional (in)dependencies revealed by the diffusion graph underlying the dependencies presented by the correlograms.
This refined picture of trait relationships might guide the search for causal links between the traits as well as help identify modularity in the evolutionary process, in which subgroups of traits are intimately connected throughout evolution.  

Another important advantage of our approach is that, with the appropriate adaptations, it can be integrated to a broad range of Gaussian models due to its readily use feature. Under a computational perspective, including graph estimation in the MBD model only requires changes in: i) the precision matrix update mechanism to jointly obtain $p(\KK, \GG)$; and ii) the prior and hyperprior choices for those parameters. Therefore, the proposed novelty does not impact the approaches employed for likelihood computations in traditional models since model's dependence on $\GG$ is completely mediated by $\KK$, i.e.\ $p(\XX|\KK,\GG,\FF)=p(\XX|\KK,\FF)$. 
This is a desirable and convenient feature because it enables our GGM approach to potentially profit from any future computational improvement in trait evolution models.
For example, in the prokaryotes application we were able to perform covariance selection on a massive data set by building upon the efficient approach developed by \citet{hassler2020}, which integrates out missing values and allows for previously intractable analyses on large trees.

The graphical model is computationally more expensive than the full model due to the additional steps required for graph estimation such as the computation of G-Wishart prior $I_G(\delta, \DD)$ and posterior $I_G(\delta+N, \DD+\bm\Delta)$ normalizing constants to perform graph updates. Additionally, chains must be run longer to account for increased complexity in the parametric space. This added computational cost is evaluated in Supplementary Material A \citep[Section A3]{pinheiro2022}. This restriction, however, is not overly limiting because when we simultaneously estimate the phylogenetic tree, the computational cost of graph estimation is small compared to the global cost of MCMC.

One possible future improvement for our approach lies in the mechanism choice to perform graph updates. Graph estimation is incredibly challenging given the dimensionality of the graph space. Our choice of graph updates using the ratio of normalizing constants is convenient because it does not require any additional implementation, since G-Wishart normalizing constant approximations should be mandatorily implemented for G-Wishart likelihood computations. In spite of convenience, this is one of the early approaches to tackle this expensive step in GGM. 
Algorithms such as birth-death MCMC (BDMCMC) \citep{mohammadi2015} are potential directions to explore. 
Additionally, in order to better characterize the performance of the presented methodology, examining a broader range of simulation conditions, such as different graph structures $\GG_0$ and trait dimension $p$, is an important future direction.

Finally, while we do not explore this in simulation or application, GPTEM can be further adapted to deal with binary data, as in \citet{zhang2021}, and/or categorical, and ordinal data as in \citet{cybis2015}, which will only add to the model's broad applicability.
The biggest challenge for this extension is how to bypass the identifiability issue on the diffusion precision. One way to achieve this is using a parameter expansion for data augmentation (PXDA) approach \citep{chib1998,talhouk2012}.

\begin{appendix}

\section*{Graph updates}\label{SUPP: graph updates}
Here we present further details to obtain the marginal distribution of $\GG$ given all the other parameters in the model, except $\KK$. Under the non-informative prior choices for diffusion graph $1/|\mathcal{G}|$ and diffusion precision $\mathcal{W}_G(3,\bf{I}_p)$, the joint density $p(\XX,\KK,\GG|\FF,\delta,\DD)$ is given by
\begin{eqnarray}\label{SUPP: joint GGM}
    p(\XX,\KK,\GG|\delta,\DD,\FF) & = & p(\XX|\KK,\FF) p(\KK|\GG,\delta,\DD,\FF) p(\GG|\delta,\DD) \nonumber \\ 
    & = & \frac{|\KK| ^{n/2} }{(2\pi)^{np/2}} \exp { \left\{-\frac{1}{2} \tr(\UU \KK ) \right\}} \nonumber\\
    &  &\quad \times
    \frac{1}{I_G(\delta,\DD)}| \KK | ^{(\delta - 2) / 2} \exp { \left\{ - \frac{1}{2} \tr(\DD \KK) \right\}}\bm{1}_{\KK \in \mathbb{P}_G}\frac{1}{|\mathcal{G}| } \nonumber\\ 
    & = &\frac{1}{(2\pi)^{np/2}}\frac{1}{|\mathcal{G}|}\frac{1}{I_G(\delta, \DD)}|\KK| ^{(\delta+N-2)/2}\exp{ \left \{ -\frac{1}{2} \tr ([\DD+\bm\Delta]\KK)  \right \}} \bm{1}_{\KK \in \mathbb{P}_G}, 
\end{eqnarray}
where $\bm \Delta =\left(\XX-\bm 1_N \bm\mu_0^t \right)^t  \left(\bm\Upsilon+\tau_0^{-1}\textbf{J}_N\right)^{-1}  \left(\XX-\bm{1}_N \bm\mu_0^t \right)$. 
In order to obtain the marginal distribution of $\XX$, given all other parameters except $\KK$, we integrate the joint distribution (\ref{SUPP: joint GGM}) over the possible values for $\KK$,
\begin{equation}\label{SUPP: marginal likelihood GGM}
     p(\XX|\GG,\delta,\DD,\FF)  = \frac{p(\XX,\GG|\delta,\DD,\FF)}{p(\GG|\delta,\DD,\FF)}=\frac{\int_{\KK  \in \mathbb{P}_G} p(\XX,\KK,\GG|\delta,\DD,\FF)\,d\KK }{p(\GG|\delta,\DD,\FF)}.
\end{equation}
By replacing the joint density (\ref{SUPP: joint GGM}) in the kernel of the integral in Equation (\ref{SUPP: marginal likelihood GGM}), we have
\begin{eqnarray}\label{SUPP: marginal likelihood X|G}
p(\XX|\GG,\delta,\DD,\FF) 
    & = &  \frac{|\mathcal{G}|}{(2\pi)^{Np/2}} \frac{1}{|\mathcal{G}|}\frac{1}{I_G(\delta, \DD)}\int_{\KK \in \mathbb{P}_G} | \KK | ^{(\delta+N-2)/2}\exp{ \left \{ -\frac{1}{2} \tr([\DD+\bm\Delta]\KK)\right \}} \,d\KK \nonumber\\
    & = & \frac{1}{(2\pi)^{Np/2}} \frac{I_G(\delta+N, \DD+\bm\Delta)}{I_G(\delta, \DD)}.
\end{eqnarray}
Note that the kernel of the integral in Equation (\ref{SUPP: marginal likelihood X|G}) corresponds to the posterior of the diffusion precision, whose distribution is $\mathcal{W}_G(\delta+N,\DD+\bm\Delta)$.
Therefore the marginal distribution of $\GG$, given all other parameters except $\KK$,
\begin{equation}\label{SUPP: posterior G|X}
    p(\GG|\XX,\delta,\DD,\FF) \propto p(\GG|\delta,\DD)p(\XX|\GG,\delta,\DD,\FF)=\frac{p(\GG|\delta,\DD)}{(2\pi)^{Np/2}} \frac{I_G(\delta+N, \DD+\bm\Delta)}{I_G(\delta, \DD)}.
\end{equation}
Hence, computing the marginal likelihood (\ref{SUPP: marginal likelihood GGM}) or the posterior distribution (\ref{SUPP: posterior G|X}) is reduced to the problem of computing normalising constants of the type $I_G(\delta, \DD)$, with $\delta>0$ and $\DD$ positive definite, which are sufficient conditions for convergence of the normalizing constants \citep[][Lemma 3.2.1: $I_G(\delta, \DD) < \infty$ for $\delta>0$]{mitsakakis2010}. 

\end{appendix}

%
%

\begin{acks}[Acknowledgments]
We thank Janira Prichula for the graph illustrations.
\end{acks}
\begin{funding}
GWH was supported through National Institutes of Health (NIH) grant F31 AI154824.
MAS was supported through NIH grants R01 AI153044 and U19 AI135995.
\end{funding}

\begin{supplement}
\stitle{A. Supplementary Sections}
\sdescription{This file contains supplementary sections that provide details regarding the simulation study, G-Wishart normalizing constant computation, graph updates and computational effort of GPTEM.}
\end{supplement}

\begin{supplement}
\stitle{B. Data set and source code}
\sdescription{We provide the Darwin's Finches and Prokaryote data sets and source codes to reproduce the application results in the article.}
\end{supplement}


\bibliographystyle{imsart-nameyear} 
\bibliography{bibliography}       




\section*{Supplementary material}

# SPARSE PRECISION MATRIX ESTIMATION IN PHYLOGENETIC TRAIT EVOLUTION MODELS (SUPPLEMENTARY MATERIAL A)

By Felipe Grillo Pinheiro[1,a], Taiane Schaedler Prass[1,b], Gabriel W Hassler[2,d], Marc A Suchard[2,3,e] and Gabriela Bettella Cybis[1,c]

[1]*Department of Statistics, Universidade Federal do Rio Grande do Sul, Porto Alegre, Brazil,* [a]*grillopf@gmail.com;* [b]*taiane.prass@ufrgs.br;* [c]*gabriela.cybis@ufrgs.br*

[2]*Department of Computational Medicine, University of California, Los Angeles, United States,* [d]*ghassler@ucla.edu*

[3]*Departments of Biostatistics and Human Genetics, University of California, Los Angeles, United States,* [e]*msuchard@ucla.edu*

**A1. Computing the G-Wishart normalizing constant.** We employ the Monte Carlo method of Atay-Kayis and Massam (2005) to compute the G-Wishart normalizing constant $I_G(\delta, \mathbf{D})$. Note that this method is valid even when the graph $\mathbf{G}$ is non-decomposable. We first state a few background definitions required to present the algorithm (for more detail, see Atay-Kayis and Massam (2005)).

Consider the notation in the Modeling Section, and let $\mathcal{V} = \{(i,j) \text{ s.t. } i = j \text{ with } i \in V \text{ or } (i,j) \in E\}$. Let a $\mathcal{V}$-incomplete matrix $\mathbf{Z}^{\mathcal{V}}$ be a $p \times p$ symmetric matrix which has entries $\mathbf{Z}_{ij}^{\mathcal{V}}$ determined if $(i,j) \in \mathcal{V}$ (diagonal element or the corresponding edge is present in graph $\mathbf{G}$), and the remainder of entries undetermined. Additionally, let $\mathbf{Z}_{\mathcal{V}}$ be the projection of $\mathbf{Z}$ onto the space of $\mathcal{V}$-incomplete matrices, then a $p \times p$ symmetric matrix $\mathbf{Z}$ is said to be a completion of $\mathbf{Z}^{\mathcal{V}}$ if $\mathbf{Z}_{\mathcal{V}} = \mathbf{Z}^{\mathcal{V}}$, that is, it coincides with $\mathbf{Z}_{\mathcal{V}}$ in all positions corresponding to edges in $\mathbf{G}$. Furthermore, if $\mathbf{Z}^{\mathcal{V}}$ is partially positive definite, then a completion $\mathbf{Z}$ such that $\mathbf{Z}^{-1} \in \mathcal{M}_G$ is unique, whenever it exists. Thus, if we know the entries of the covariance matrix $\mathbf{\Sigma}$ that coincide with the edges of $\mathbf{G}$, then the remaining entries are a deterministic function of the known entries, if the precision $\mathbf{K}$ is to be positive definite and have zero entries whenever $(i,j) \notin \mathcal{V}$. The completion formula (for the Cholesky decomposition of $\mathbf{Z}^{\mathcal{V}}$) can be found in Proposition 2 of Atay-Kayis and Massam (2005).

The Monte Carlo method is justified by applying two variable transformations to the target integral

$$(A1.1) \qquad I_G(\delta, \mathbf{D}) = \int_{\mathbf{K} \in \mathbb{P}_G} |\mathbf{K}|^{(\delta-2)/2} \exp\left\{-\frac{1}{2}\mathrm{tr}(\mathbf{DK})\right\} d\mathbf{K}.$$

First the Cholesky transform is applied to the precision matrix, $\mathbf{K} = \phi^t \phi$ with $\phi$ upper triangular, and $\mathbf{K}$ is mapped to the $\mathcal{V}$-incomplete upper triangular matrix $\phi^{\mathcal{V}}$. Then a second variable transform is applied in which $\phi^{\mathcal{V}}$ is mapped to $\psi^{\mathcal{V}}$, with $\psi = \phi \mathbf{T}^{-1}$, where $\mathbf{T}$ comes from the Cholesky decomposition of the relevant rate matrix $\mathbf{D} = (\mathbf{T}^t \mathbf{T})^{-1}$. This yields the kernel of a Wishart distribution with identity rate matrix. Then, exploring the properties of this distribution, the target integral from expression (A1.1) can be seen as a constant multiplied by the expectation of a function of a series of independent univariate Gaussian and Chi-squared distributions. That is,

$$(A1.2) \qquad I_G(\delta, \mathbf{D}) = C_{\delta, \mathbf{D}} \cdot \mathbb{E}\left[f_T(\psi^{\mathcal{V}})\right],$$

in which the constant $C_{\delta, \mathbf{D}}$ is defined in equation (45) of Atay-Kayis and Massam (2005) and

$$(A1.3) \qquad f_T(\psi^{\mathcal{V}}) = \exp\left\{-\frac{1}{2}\sum_{(i,j)\notin\mathcal{V}}\psi_{ij}^2\right\}.$$

The expectation is such that

$$(A1.4) \qquad \psi_{ii}^{\mathcal{V}} \sim \sqrt{\chi_{\delta+\nu_i}^2}, \quad i = 1, \cdots, p,$$

with $\nu_i$ representing the degree of edge $i$ in $\mathbf{G}$, and

$$(A1.5) \qquad \psi_{ij}^{\mathcal{V}} \sim \mathcal{N}(0,1), \quad i \neq j \in \mathcal{V}.$$

The random variables above are pairwise independent. Note that $f_T(\psi^{\mathcal{V}})$ is in fact a deterministic function of $\psi^{\mathcal{V}}$, since the entries of $\psi$ not in $\mathbf{G}$ are uniquely determined by the completion formula. Thus, the Monte Carlo method based on this formulation can be expressed as follows

1. For replication $r \in \{1, \cdots, Re\}$
   a) Sample the $\psi_{ii,r}$ from independent $\sqrt{\chi_{\delta+\nu_i}^2}$, for $i = 1, \cdots, p$,
   b) Sample the $\psi_{ij,r}$ from independent $\mathcal{N}(0,1)$, for $(i,j) \in \mathcal{V}$,
   c) Compute the $\psi_{ij,r}$ not in graph from the completion formula
2. Estimate $I_G(\delta, \mathbf{D}) = C_{\delta, \mathbf{D}} \cdot \frac{1}{Re} \sum_{r=1}^{Re} f_T(\psi_r^{\mathcal{V}})$

**A2. Simulation Study.** Here we provide details about the simulations conducted to compare the performances of both graphical and full models. We first present the true diffusion precision matrices and the corresponding diffusion correlation matrices for simulations $Sim \ 1$ and $Sim \ 2$.

For $Sim \ 1$ we define $\mathbf{K}_0$ and $\mathbf{R}_0$ as

$$(A2.1) \qquad \mathbf{K}_0 := \begin{pmatrix} 3.87 & 3.62 & 0 & 0 & 0 \\ 3.62 & 8.50 & -4.87 & 0 & 0 \\ 0 & -4.87 & 5.13 & 0 & 0 \\ 0 & 0 & 0 & 6.55 & -6.29 \\ 0 & 0 & 0 & 6.29 & 6.55 \end{pmatrix},$$

and

$$(A2.2) \qquad \mathbf{R}_0 := \begin{pmatrix} 1.00 & -0.93 & -0.89 & 0 & 0 \\ -0.93 & 1.00 & 0.95 & 0 & 0 \\ -0.89 & 0.95 & 1.00 & 0 & 0 \\ 0 & 0 & 0 & 1.00 & 0.96 \\ 0 & 0 & 0 & 0.96 & 1.00 \end{pmatrix}$$

For $Sim \ 2$ we define

$$(A2.3) \qquad \mathbf{K}_0 := \begin{pmatrix} 9.52 & -7.26 & 4.25 & -3.64 & 0 & 0 & 0 & 0 & 0 & 0 \\ -7.26 & 8.17 & -3.55 & 2.34 & 0 & 0 & 0 & 0 & 0 & 0 \\ 4.25 & -3.55 & 10.3 & 3.34 & 0 & 0 & 0 & 0 & 0 & 0 \\ -3.64 & 2.34 & 3.34 & 4.46 & 0 & 0 & 0 & 0 & 0 & 0 \\ 0 & 0 & 0 & 0 & 0.59 & -0.59 & 0 & 0 & 0 & 0 \\ 0 & 0 & 0 & 0 & -0.59 & 2.91 & -1.54 & 0 & 0 & 0 \\ 0 & 0 & 0 & 0 & 0 & -1.54 & 1.35 & 0 & 0 & 0 \\ 0 & 0 & 0 & 0 & 0 & 0 & 0 & 7.84 & 1.12 & 0.12 \\ 0 & 0 & 0 & 0 & 0 & 0 & 0 & 1.12 & 2.86 & 1.18 \\ 0 & 0 & 0 & 0 & 0 & 0 & 0 & 0.12 & 1.18 & 3.16 \end{pmatrix},$$

and

$$(A2.4) \quad \mathbf{R}_0 := \begin{pmatrix} 1 & 0.6 & -0.96 & 0.97 & 0 & 0 & 0 & 0 & 0 & 0 \\ 0.6 & 1 & -0.41 & 0.43 & 0 & 0 & 0 & 0 & 0 & 0 \\ -0.96 & -0.41 & 1 & -0.98 & 0 & 0 & 0 & 0 & 0 & 0 \\ 0.97 & 0.43 & -0.98 & 1 & 0 & 0 & 0 & 0 & 0 & 0 \\ 0 & 0 & 0 & 0 & 1 & 0.72 & 0.62 & 0 & 0 & 0 \\ 0 & 0 & 0 & 0 & 0.72 & 1 & 0.87 & 0 & 0 & 0 \\ 0 & 0 & 0 & 0 & 0.62 & 0.87 & 1 & 0 & 0 & 0 \\ 0 & 0 & 0 & 0 & 0 & 0 & 0 & 1 & -0.25 & 0.08 \\ 0 & 0 & 0 & 0 & 0 & 0 & 0 & -0.25 & 1 & -0.4 \\ 0 & 0 & 0 & 0 & 0 & 0 & 0 & 0.08 & -0.4 & 1 \end{pmatrix}.$$

Moreover, Table A2.1 shows the definition of the statistics used to evaluate the performances of both full and graphical models presented in Simulation Study Section based on the confusion matrix.

TABLE A2.1

*Definition of the statistics used to assess the performances of graph estimation in the graphical model and correlation selection with HPD criteria in the full model. TP = True Positive; TN = True Negative; FP = False Positive; FN = False Negative.*

| Statistic | Definition |
|---|---|
| Accuracy | $\dfrac{TP + TN}{TP + FP + FN + TN}$ |
| Sensitivity | $\dfrac{TP}{TP + FN}$ |
| Specificity | $\dfrac{TN}{TN + FP}$ |
| Precision | $\dfrac{TP}{TP + FP}$ |
| F1-Score | $\dfrac{2TP}{2TP + FP + FN}$ |

**A3. Computational Effort.** Here we provide a summary of the computational cost of including graph estimation in GPTEM. We formalize our comparison by computing the parameter update cost, i.e. the proportion of time spent updating the parameter of interest ($\mathbf{K}$ in full model and $(\mathbf{K}, \mathbf{G})$ in graphical model), and the minimum and median effective sample size (ESS) per minute for the diffusion correlation under both graphical and full models (Table A3.1). We fixed the phylogenetic tree in the simulation study while estimate $\mathscr{F}$ in the application data sets. Notice that, in the simulations, we can directly compare the computational cost of equipping the MBD model with GGM because we update $(\mathbf{K}, \mathbf{G})$ in the graphical model, but only $\mathbf{K}$ in the full model. This is the reason why parameter update cost is 100% in both $Sim$ 1 and $Sim$ 2. When the tree $\mathscr{F}$ is fixed, we see that ESS per second decreases between 76 and 82%. Indeed, graphical models are computationally more expensive compared to full models due to the graph update burden.

On the other hand, when the tree is simultaneously estimated, most of the computational effort involves the phylogenetic tree estimation such that the computational cost of structure learning decreases compared to the global MCMC cost. One can see that the computational cost of updating the diffusion graph is less than 5% in the worst scenario. The Darwin's

<div style="text-align:center">

TABLE A3.1

*Computational cost of graph estimation. We report MCMC sampling efficiency through the update parameter cost, which is the proportion of time spent to update the parameter of interest in comparison to global MCMC cost, and by computing the minimum and median effective sample size (ESS) per minute.*

</div>

| Data set | $\mathscr{F}$ | $N$ | $p$ | Model | Update parameter | cost | $\mathbf{R}$ ESS/minute minimum | median |
|---|---|---|---|---|---|---|---|---|
| *Sim 1* | Fixed | 50 | 5 | Full | $\mathbf{K}$ | 100% | 4445.59 | 5069.41 |
| | | | | Graphical | $(\mathbf{K}, \mathbf{G})$ | 100% | 1053.96 | 1189.19 |
| | | | | **Speed-down** | - | - | **76.3%** | **76.5%** |
| *Sim 2* | Fixed | 100 | 10 | Full | $\mathbf{K}$ | 100% | 1606.49 | 2273.61 |
| | | | | Graphical | $(\mathbf{K}, \mathbf{G})$ | 100% | 285.94 | 426.14 |
| | | | | **Speed-down** | - | - | **82.2%** | **81.3%** |
| *Darwin's finches* | Estimated | 13 | 5 | Full | $\mathbf{K}$ | 0.66% | 1244.43 | 1244.43 |
| | | | | Graphical | $(\mathbf{K}, \mathbf{G})$ | 4.83% | 887.56 | 992.03 |
| | | | | **Speed-down** | - | - | **28.7%** | **20.3%** |
| *Prokaryotes* | Estimated | 705 | 7 | Full | $\mathbf{K}$ | 1.26% | 1.02 | 2.45 |
| | | | | Graphical | $(\mathbf{K}, \mathbf{G})$ | 1.30% | 0.35 | 1.84 |
| | | | | **Speed-down** | - | - | **66.1%** | **24.8%** |

Finches and $Sim\ 1$ are both $p = 5$ data sets, but the speed-down for the median ESS/minute drops from 76.5% in $Sim\ 1$ to 20.3% in finches data. Note also that in the prokaryotes application, although updating $(\mathbf{K}, \mathbf{G})$ requires more computational effort than updating $\mathbf{K}$ only (full model), both update costs are negligible compared to the global update cost of MCMC (1.26% and 1.30%, respectively).

<div style="text-align:center">

REFERENCES

</div>


\end{document}